\documentclass[aps,prb,twocolumn,superscriptaddress,10pt,floatfix]{revtex4-2}
\usepackage[colorlinks=true,citecolor=blue,linkcolor=blue,urlcolor=blue,]{hyperref}
\usepackage{graphicx,nicefrac,textcomp}
\usepackage{float}
\usepackage[normalem]{ulem}
\usepackage{amssymb,amsmath}
\usepackage{soul}
\usepackage{color}
\usepackage{textcomp}
\DeclareRobustCommand{\textendash}{\ifmmode\text{--}\else\leavevmode\hbox{--}\fi}
\usepackage{gensymb}
\usepackage{graphics}\usepackage{subfigure}\usepackage{pstricks}\usepackage{dcolumn}\usepackage{bm}
\usepackage{booktabs}
\usepackage{multirow}
\usepackage{siunitx}

\begin{document}

\title{High temperature transitions in Ruddlesden-Popper nickelates La$_{n+1}$Ni$_{n}$O$_{3n+1}$}
\author{P.~Reiss}
\affiliation{Max Planck Institute for Solid State Research, Heisenbergstra{\ss}e 1, 70569 Stuttgart, Germany}
\author{A.~Shevchenko}
\affiliation{Max Planck Institute for Solid State Research, Heisenbergstra{\ss}e 1, 70569 Stuttgart, Germany}
\author{P. S. Lizama}
\affiliation{Max Planck Institute for Solid State Research, Heisenbergstra{\ss}e 1, 70569 Stuttgart, Germany}
\author{J. Nuss}
\affiliation{Max Planck Institute for Solid State Research, Heisenbergstra{\ss}e 1, 70569 Stuttgart, Germany}
\author{R. Dinnebier}
\affiliation{Max Planck Institute for Solid State Research, Heisenbergstra{\ss}e 1, 70569 Stuttgart, Germany}
\author{P.~A.~van~Aken}
\affiliation{Max Planck Institute for Solid State Research, Heisenbergstra{\ss}e 1, 70569 Stuttgart, Germany}
\author{M.~Hepting}
\affiliation{Max Planck Institute for Solid State Research, Heisenbergstra{\ss}e 1, 70569 Stuttgart, Germany}
\author{M.~Isobe}
\affiliation{Max Planck Institute for Solid State Research, Heisenbergstra{\ss}e 1, 70569 Stuttgart, Germany}
\author{Y.~E.~Suyolcu}
\affiliation{Max Planck Institute for Solid State Research, Heisenbergstra{\ss}e 1, 70569 Stuttgart, Germany}
\author{H. Takagi}
\affiliation{Max Planck Institute for Solid State Research, Heisenbergstra{\ss}e 1, 70569 Stuttgart, Germany}
\author{B.~Keimer}
\affiliation{Max Planck Institute for Solid State Research, Heisenbergstra{\ss}e 1, 70569 Stuttgart, Germany}
\author{P.~Puphal}
\email{puphal@fkf.mpg.de}
\affiliation{Max Planck Institute for Solid State Research, Heisenbergstra{\ss}e 1, 70569 Stuttgart, Germany}

\date{\today}

\begin{abstract}
The discovery of superconductivity at $15\,\mathrm{K}$ in the infinite-layer nickelate $(\mathrm{Nd},\mathrm{Sr})\mathrm{NiO}_2$, followed by superconductivity at $80\,\mathrm{K}$ in the Ruddlesden--Popper phase $\mathrm{La}_3\mathrm{Ni}_2\mathrm{O}_7$, has ushered in a new era of nickelate research. Despite this progress, large discrepancies between reports exist. Here, we investigate the complete series of bulk-stable $\mathrm{La}_{n+1}\mathrm{Ni}_n\mathrm{O}_{3n+1}$ compounds using a comprehensive set of experimental techniques, including PXRD, single-crystal XRD, electron microscopy, heat capacity, differential scanning calorimetry, magnetic susceptibility, and transport measurements, over a broad temperature range from $2$ to $1000\,\mathrm{K}$. By studying high-quality single crystals, we identify a previously underappreciated high-temperature phase transition in Ruddlesden--Popper nickelates $\mathrm{La}_{n+1}\mathrm{Ni}_n\mathrm{O}_{3n+1}$ distinct from the one going to a tetragonal phase. 
\end{abstract} 

\maketitle
\footnotetext[1]{See Supplemental Material at [URL will be inserted by publisher] for further details, which includes Refs xxx.} 

\section{Introduction}
Ruddlesden--Popper (RP) nickelates with the general formula $RE_{n+1}$Ni$_n$O$_{3n+1}$ ($RE$ = rare-earth ion) have attracted sustained interest as potential analogues to cuprate superconductors because of their layered crystal structure and correlated electronic behavior \cite{Greenblatt1997}. The idea that RP nickelates could host superconductivity predates the discovery of high-$T_c$ cuprates \cite{Bednorz1988}. Early theoretical proposals focused on the $n=1$ member, $RE_2$NiO$_4$, motivated by its structural similarity to La$_2$CuO$_4$ and the observation of charge and spin stripe order \cite{Acrivos1994,Sachan1995}, phenomena that are closely associated with superconductivity in hole-doped cuprates.

Despite these early expectations, superconductivity was not observed in $n=1$ RP nickelates, a failure commonly attributed to their formal Ni$^{2+}$ ($3d^8$) configuration with spin $S=1$ \cite{Sugai1990}, which differs fundamentally from the $3d^9$ ($S=1/2$) configuration of Cu$^{2+}$ in cuprates. Higher-order RP members ($n \ge 2$) interpolate between this insulating limit and the perovskite nickelates $RE$NiO$_3$ ($n=\infty$), and exhibit increasing three-dimensionality and mixed-valence Ni states. In contrast to infinite-layer nickelates, RP systems retain strongly coupled NiO$_6$ octahedral layers, leading to a multi-orbital low-energy electronic structure in which both Ni $3d_{x^2-y^2}$ and $3d_{z^2}$ states contribute at the Fermi level \cite{Sun2023,Luo2023,Zhang2023b,Wang2024b}.

A major breakthrough occurred with the 2023 discovery of superconductivity in bilayer La$_3$Ni$_2$O$_7$ under pressures exceeding 14~GPa, where transition temperatures as high as $T_c \sim 80$~K were reported \cite{Sun2023}. This finding motivated the observation of superconductivity in other RP nickelates, including trilayer La$_4$Ni$_3$O$_{10}$ with $T_c$ values of 20--30~K \cite{Sakakibara2024,Zhu2024}, as well as in hybrid RP structures composed of stacked mono-, bi-, and trilayer blocks \cite{Chen2024,Puphal2024,Wang2024a,Li2024,Shi2025,Huang2025}. In addition to hydrostatic pressure, epitaxial strain and rare-earth substitution have been shown to enhance superconductivity, with compressively strained La$_3$Ni$_2$O$_7$ thin films reaching $T_c \sim 40$~K and partial Pr substitution improving reproducibility and transition sharpness \cite{Ko2024,Zhou2025F,Liu2025}.

Overall, RP nickelates define a distinct family of superconductors characterized by intermediate Ni valence states ranging from $3d^{7.33}$ in La$_4$Ni$_3$O$_{10}$ to $3d^{7.5}$ in La$_3$Ni$_2$O$_7$, strong interlayer coupling, and multiorbital electronic character. Their superconducting phenomenology differs markedly from both cuprates and infinite-layer nickelates, with the highest transition temperatures observed in bilayer compounds rather than trilayers. These discoveries establish RP nickelates as a new platform for exploring unconventional superconductivity and underscore the importance of structural dimensionality, pressure, and synthesis control in shaping their electronic ground-states \cite{Sun2023,Puphal2024,Puphal2025}.

Reproducibility remains a major challenge in RP nickelates. Reports indicate inhomogeneous \cite{Mandyam2026}, or filamentary superconductivity \cite{Puphal2024}. A pronounced sample dependence is observed \cite{Sun2023,Puphal2024,Chen2024,Wang2024a}, affecting not only the stabilized structural phase but also the realized oxygen content \cite{Dong2024,Dong2025} and, consequently, the measured physical properties. Because superconductivity emerges only at low temperatures, recent studies have naturally focused on the ground state and its properties. However, long before the discovery of superconductivity, high-temperature anomalies were already reported in polycrystalline RP nickelates, indicating that essential aspects of their physics extend well beyond the low-temperature regime \cite{Kobayashi1996,AMOW2006,Zinkevich2007,Sasaki1997,Tamura1996,Hayashi1993}.

Here, we systematically investigate the complete series of bulk-stable single crystals of RP La$_{n+1}$Ni$_n$O$_{3n+1}$ using a broad set of experimental techniques over an extended temperature range from 2 to 1000~K, with the aim of confirming these high-temperature anomalies in crystals of $n=1,2,3$ and its absence in $n=\infty$.

\section{Results}
\subsection{RP La$_{n+1}$Ni$_n$O$_{3n+1}$ phases}
\begin{figure}[h]
    \centering
    \includegraphics[width=1\linewidth]{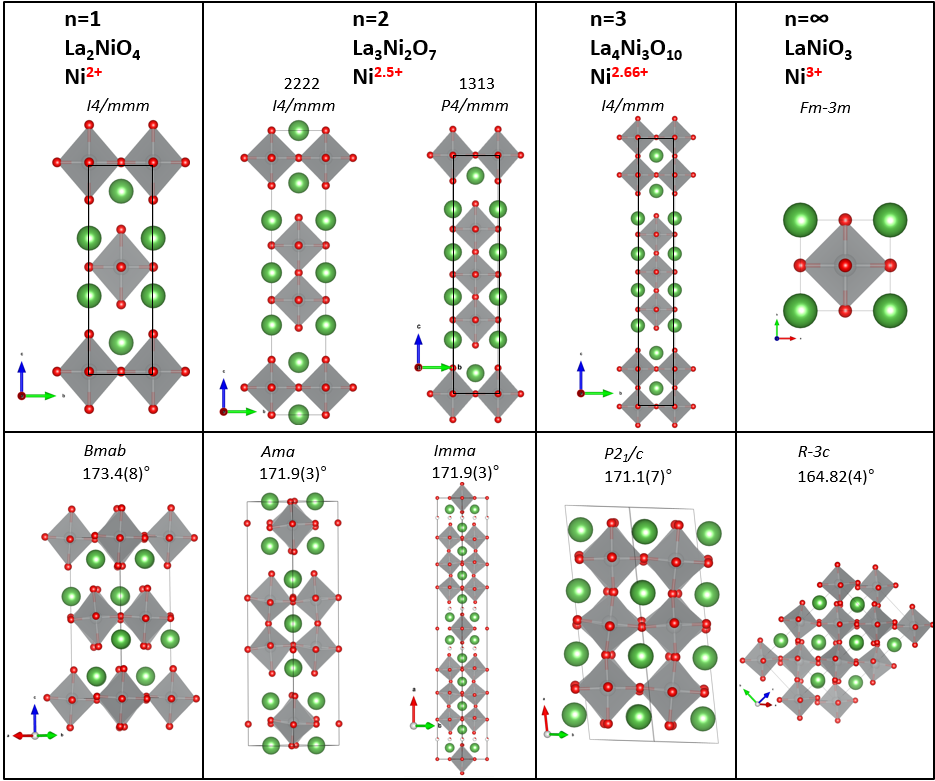}
    \caption{Crystal structures of Ruddlesden--Popper nickelates $\mathrm{La}_{n+1}\mathrm{Ni}_n\mathrm{O}_{3n+1}$ for $n=1,2,3$, and $\infty$. The first row shows the high-temperature tetragonal or cubic (HTT) structures, while the second row displays the corresponding low-temperature buckled variants. The space group of each phase is indicated in the figure. Color code: La (green), Ni (grey polyhedra), and O (red).}
    \label{struc}
\end{figure}

The Ruddlesden--Popper (RP) structure constitutes a homologous series of compounds with the general formula $A_{n+1}B_nX_{3n+1}$, in which $n$ layers of corner-sharing perovskite units are periodically interrupted by a rocksalt-type layer, as illustrated in Fig.~\ref{struc}. Shown are the $n=1$ member $\mathrm{La}_2\mathrm{NiO}_4$, the $n=2$ compound $\mathrm{La}_3\mathrm{Ni}_2\mathrm{O}_7$ in its two polymorphic realizations (bilayer ``2222'' and monolayer--trilayer ``1313''), the $n=3$ phase $\mathrm{La}_4\mathrm{Ni}_3\mathrm{O}_{10}$, and the perovskite end member $n=\infty$, $\mathrm{LaNiO}_3$.  

As is evident from the chemical formula, the $A$ site in nickelate RP phases is conventionally occupied by a trivalent rare-earth ion ($A=RE$). To date, the largest range of bulk-stable compounds with $n=1,2,3$, and $\infty$ have been reported exclusively for $RE=\mathrm{La}$. Figure~\ref{struc} presents these structures in the first row in their high-temperature tetragonal variants, while the second row depicts the corresponding low-temperature buckled structures.  Comparing the room-temperature structures from literature, the Ni-O-Ni bond angles  monotonically decrease with $n$, with representative examples moving from $173.4(8)^\circ$, $171.9(3)^\circ$, $171.1(7)^\circ$, and $164.8(4)^\circ$ for $n=1,2,3$, and $\infty$, respectively, corresponding to increasing octahedral buckling.

\subsection{PXRD}

\begin{figure}[h]
    \centering
    \includegraphics[width=1\linewidth]{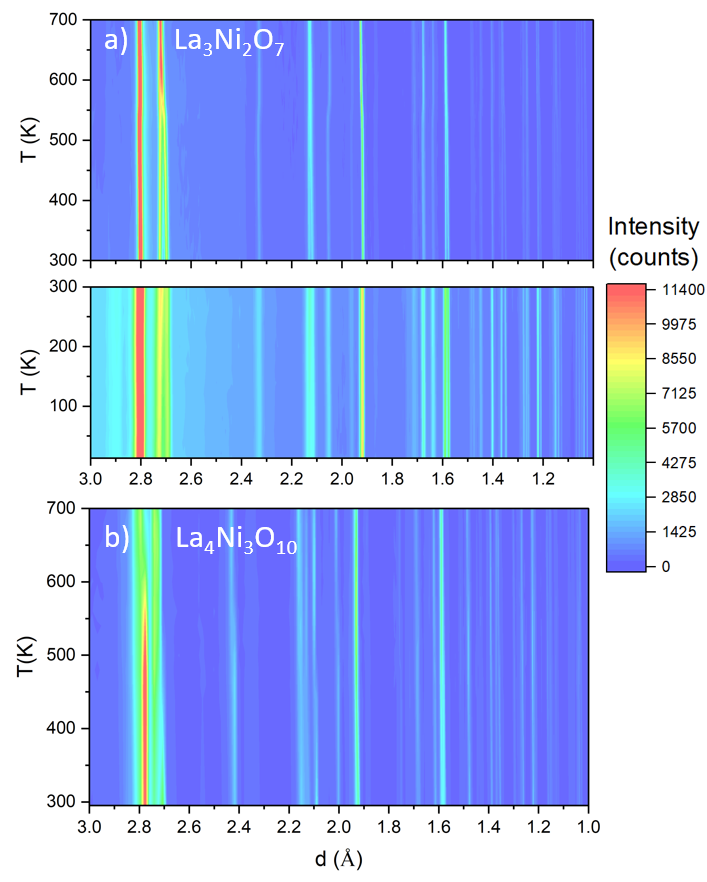}
    \caption{Contour plots of powder X-ray diffraction (PXRD) data for Ruddlesden-Popper nickelates $\mathrm{La}_{n+1}\mathrm{Ni}_n\mathrm{O}_{3n+1}$ for $n=2,3$. The plots show intensity (color scale) as a function of temperature (y-axis) and d-spacing (x-axis).}
    \label{contour}
\end{figure}

\begin{figure*}
    \centering
    \includegraphics[width=1\linewidth]{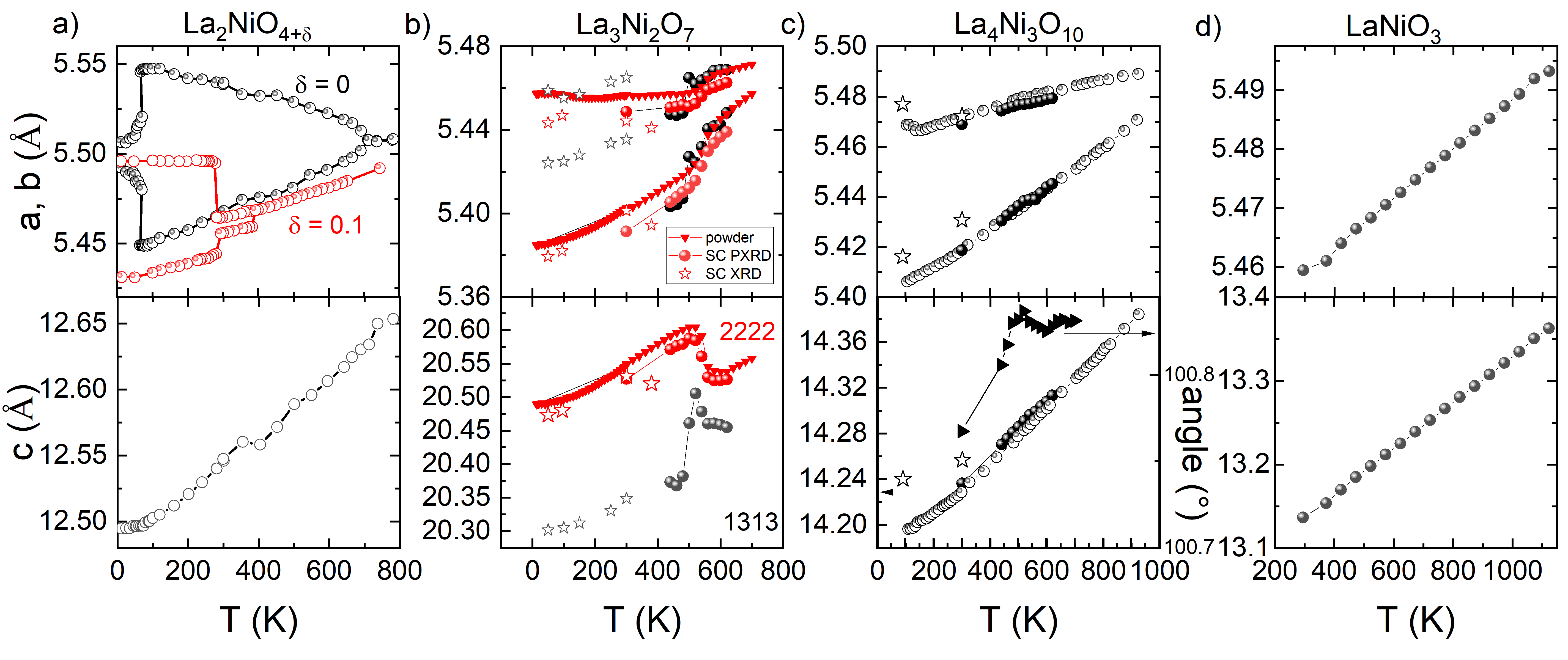}
    \caption{Temperature evolution of lattice parameters extracted from Rietveld refinement of PXRD data (see supplemental Materials) and from single-crystal XRD (see Fig. \ref{scXRD} and \ref{LTscXRD}) for Ruddlesden--Popper $\mathrm{La}_{n+1}\mathrm{Ni}_n\mathrm{O}_{3n+1}$: (a) $n=1$, showing both $\delta=0$ reduced samples (black) and $\delta=0.1$ as-grown samples (red, digitized from Ref.~\cite{Tamura1996}); (b) $n=2$ in the two polymorphic forms, bilayer 2222 (red) and monolayer--trilayer 1313 (black); (c) $n=3$ including its monoclinic angle on the right axis; and (d) the perovskite end member $n=\infty$.}
    \label{lattice}
\end{figure*}

We employed powder X-ray diffraction (PXRD) to investigate the RP nickelates from $n=1$ to $n=\infty$ in varying ranges of  $13$--$700~\mathrm{K}$ and 300-1075 K. In Fig. \ref{contour} we show the contour plots of $n=2,3$, where a structural anomaly at around 560 K becomes evident from sudden intensity changes and peak shifts. This transition is independent of the tetragonal-orthorhombic/monoclinic transition, as will be discussed in the following. Using Rietveld refinement of the data shown in more detail in the supplemental Material, we extract the temperature dependence of the lattice constants and summarize the results in Fig.~\ref{lattice}.

For the $n=1$ compound $\mathrm{La}_2\mathrm{NiO}_4$ [Fig.~\ref{lattice}(a), data digitized from Ref.~\cite{Tamura1996}], it is well established that upon cooling the system undergoes a second-order structural transition from tetragonal $I4/mmm$ to orthorhombic $Bmab$ at approximately $680~\mathrm{K}$, at which point the lattice parameters $a$ and $b$ begin to split. At $80~\mathrm{K}$, a magnetic transition drives these lattice parameters back together. In addition, a more subtle transition near $350~\mathrm{K}$ produces a small discontinuity in $c$ and a slight deviation between $a$ and $b$. Interestingly, oxygen intercalation to $\delta=0.1$ substantially modifies the high-temperature structural behavior: the transition is shifted down to about $390~\mathrm{K}$ and becomes first order, leading to a transition into an $Fmmm$ phase. In this case, only a weak initial splitting of $a$ and $b$ is observed, followed by a more pronounced separation near $300~\mathrm{K}$, where the system becomes more strongly orthorhombic without developing octahedral buckling.  

We next consider the RP $n=2$ compound $\mathrm{La}_3\mathrm{Ni}_2\mathrm{O}_7$, whose lattice-parameter evolution is shown in Fig.~\ref{lattice}(b). The figure combines data from crushed single crystals measured by PXRD (squares), polycrystalline samples synthesized by solid-state sintering (triangles), and single-crystal XRD measurements (stars). For the bilayer 2222 phase (red), powder and single-crystal data are in excellent agreement, aside from a slight volume contraction in the single crystals. This agreement indicates that, for selected studies, high-quality powder samples can reliably capture the intrinsic structural evolution, as discussed further below. A sharp anomaly is observed near $560~\mathrm{K}$, which does not correspond to the conventional tetragonal-to-orthorhombic transition, as the lattice parameters $a$ and $b$ remain split across this temperature. Instead, this transition is characterized by an increase of the out-of-plane lattice parameter and a concomitant decrease of the in-plane parameters, indication a sudden release of octahedral tilting. Notably, this transition was already reported in powder studies in Ref.~\cite{Sasaki1997}, but it has received little attention in more recent work focusing on superconductivity in nickelates. By contrast, the widely discussed density-wave transition near $150~\mathrm{K}$ produces only a very small increase in the $a$ lattice parameter.  

The monolayer--trilayer 1313 polymorph exhibits an anomaly at a similar temperature, manifested as a sudden volume contraction in which all lattice parameters decrease simultaneously. The slightly larger lattice constants obtained from single-crystal XRD reflect the oxygen-intercalated nature of this measured crystal. A linear extrapolation of the high-temperature behavior of $a$ and $b$ suggests that these parameters would converge at the highest measured temperatures and that oxygen intercalation may suppress the transition, similarly to the behavior observed in La$_2$NiO$_{4+\delta}$.  

Figure~\ref{lattice}(c) displays the lattice parameters of the trilayer $n=3$ compound $\mathrm{La}_4\mathrm{Ni}_3\mathrm{O}_{10}$, together with the monoclinic angle $\beta$. The overall behavior is dominated by a nearly linear thermal contraction, interrupted by a subtle anomaly near $560~\mathrm{K}$. This feature is more pronounced in $b$ than in $a$ and is also visible in $c$, but it is most clearly resolved in the temperature dependence of the monoclinic angle $\beta$. As in the $n=2$ case, the well-known low-temperature density-wave transition induces only a very small increase in the $a$ lattice parameter. This transition has attracted significant recent interest, particularly in pressure studies that explore its suppression \cite{Zhu2024}. Interestingly, the $\sim560~\mathrm{K}$ anomaly was previously detected in thermal expansion measurements \cite{Zhang2020M}, but not directly in lattice-parameter refinements.  

Finally, Fig.~\ref{lattice}(d) shows the temperature dependence of the lattice parameters of the perovskite end member $\mathrm{LaNiO}_3$. For this metallic Ni$^{3+}$ compound, no structural transitions are observed up to the highest measured temperatures of approximately $1150~\mathrm{K}$. In our crushed single crystals, we do not observe a transition to a cubic phase yet, which has been proposed to occur near $1100~\mathrm{K}$ \cite{Medarde1997}. Instead, a linear extrapolation of the splitting between the $(104)$ and $(2\bar{1}0)$ reflections suggests that the cubic transition in these crystals would occur above approximately $1400~\mathrm{K}$.

Taken together, these results reveal a rich hierarchy of structural anomalies across all layered RP nickelates, highlighting the need for a systematic and unified evaluation of their temperature-dependent structural behavior.

\subsection{Thermal studies}

\begin{figure}
    \centering
    \includegraphics[width=1\linewidth]{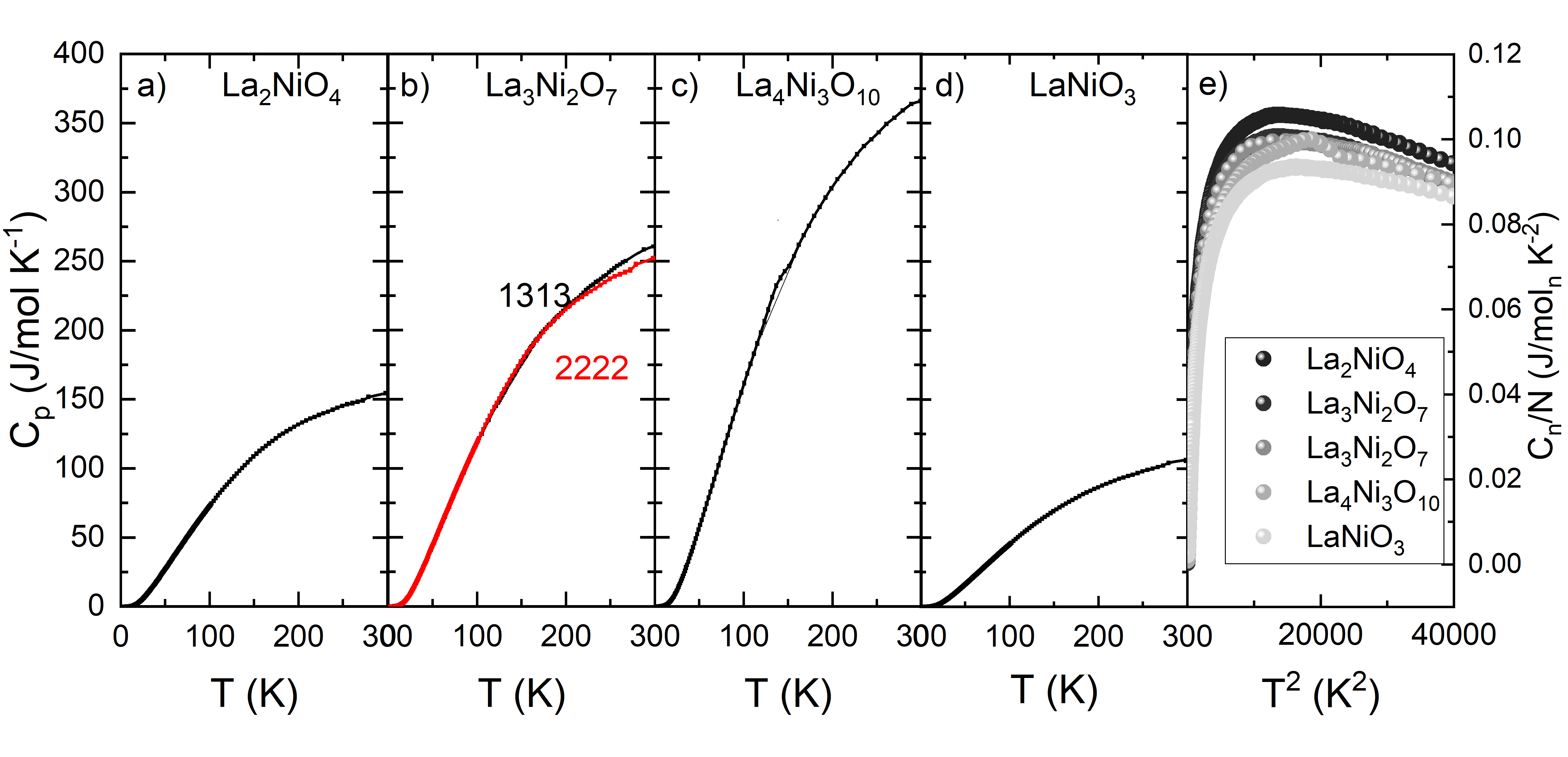}
    \caption{Specific heat $C_p$ of Ruddlesden--Popper $\mathrm{La}_{n+1}\mathrm{Ni}_n\mathrm{O}_{3n+1}$ for (a) a single crystal with $n=1$, (b) a single crystal of $n=2$ (1313) together with powder data for the 2222 polymorph (red), (c) a single crystal with $n=3$, and (d) a single crystal of the perovskite end member $n=\infty$. (e) Comparative plot of normalized $C_p/T$.}
    \label{Cp}
\end{figure}

As a next step, we performed specific-heat measurements over $2$--$300~\mathrm{K}$ on  representative samples for each value of $n$, as shown in Fig.~\ref{Cp}. With the exception of the $\mathrm{La}_3\mathrm{Ni}_2\mathrm{O}_7$ variant, most of these data have been reported previously. A direct comparison of panels (a)--(d), which share a common $y$ axis, highlights the systematic increase of the molar specific heat $C_p$ with the number of atoms per formula unit N, namely $7$, $12$, $17$, and $5$ for $n=1,2,3$, and $n=\infty$ respectively. Notably, for $\mathrm{La}_3\mathrm{Ni}_2\mathrm{O}_7$--2222 we employed powder samples, as the heat-capacity signal of available single crystals was too small. This may account for the slightly reduced high-temperature $C_p$ values, potentially arising from a small amorphous fraction or a minor oxygen deficiency. Overall, the specific-heat curves are broadly similar across the series, with only selected low-temperature anomalies.

To emphasize the systematic trends, Fig.~\ref{Cp}(e) presents a normalized comparison in which $C_p/T$ is divided by the number of atoms per formula unit. This representation reveals a clear overlap of the low-temperature linear regimes described by
$
\frac{C_p}{T} = \gamma + \beta T^2,
$
corresponding to
$
C_p = \gamma T + \beta T^3 .
$
At higher temperatures, deviations emerge, with a gradual reduction of $C_p/T$ when progressing from smaller to larger $n$, concomitant with the evolution from insulating to metallic ground states. The coefficient $\beta \propto N/\Theta_D^3$ indicates that the RP members have comparable average lattice stiffness and similar Debye temperatures.

From linear fits in the temperature range $8$--$36~\mathrm{K}$, we extract Debye temperatures of $\Theta_D \approx 340~\mathrm{K}$, $347~\mathrm{K}$, $358~\mathrm{K}$, and $367~\mathrm{K}$ for $n=1$, $n=2$, $n=3$, and $n=\infty$, respectively. These values reflect an overall similarity in lattice stiffness across the series, with a subtle increase of $\Theta_D$ with increasing $n$, in good agreement with reported values \cite{Wu2001} and their known sensitivity to oxygen stoichiometry \cite{Taniguchi1995}.

The density-wave (DW) transition in $\mathrm{La}_4\mathrm{Ni}_3\mathrm{O}_{10}$ produces a pronounced and sufficiently sharp entropy release to be clearly resolved in the specific heat. The entropy within amounts to 0.427 J/mol K$^{-1}$ compared to Ni$^{3+}$ with S = 1/2 yielding for full order Rln2 = 5.76 J mol$^{-1}$ K$^{-1}$ and Ni$^{2+}$ with S = 1 $\Delta S=$Rln3 = 9.13 J mol$^{-1}$ K$^{-1}$, so overall only 5 \% of the expected entropy for full magnetic order is released. For $\mathrm{La}_3\mathrm{Ni}_2\mathrm{O}_7$ the entropy change associated with the magnetic transition is even more subtle staying below 0.1 J/mol and becomes discernible only in a $C_p/T$ versus $T^2$ representation, where a weak anomaly appears around $T^2 \approx 2.6 \times 10^4~\mathrm{K}^2$, corresponding to $T \approx 161~\mathrm{K}$.

\begin{figure}
    \centering
    \includegraphics[width=1\linewidth]{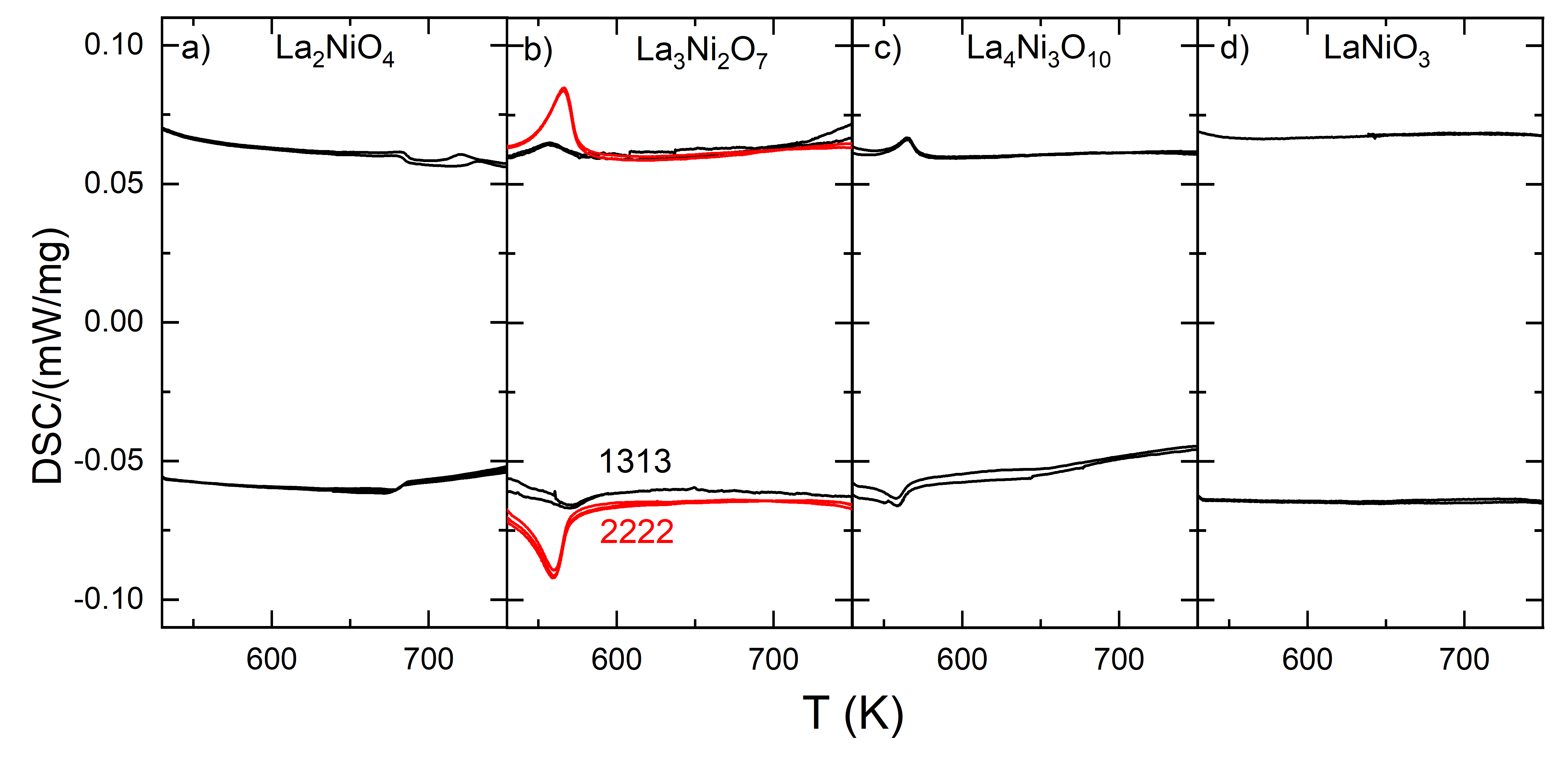}
    \caption{Differential scanning calorimetry (DSC) measurements of Ruddlesden--Popper $\mathrm{La}_{n+1}\mathrm{Ni}_n\mathrm{O}_{3n+1}$ for (a) $n=1$, (b) $n=2$ with 1313 (black) and 2222 (red) polymorphs, (c) $n=3$, and (d) $n=\infty$.}
    \label{DSC}
\end{figure}

While specific-heat measurements provide detailed insight at low temperatures, the investigation of high-temperature transitions is more effectively carried out using differential scanning calorimetry (DSC). Figure~\ref{DSC} shows DSC traces for the same set of samples over a selected temperature window of $530$--$750~\mathrm{K}$, encompassing the anomalies observed in PXRD. In Fig.~\ref{DSC}(a), $\mathrm{La}_2\mathrm{NiO}_4$ exhibits a clear transition at $680~\mathrm{K}$, along with a weaker, cycle-dependent feature at $720$--$730~\mathrm{K}$, likely caused by slight oxidation due to residual air in the pinch-sealed Al crucible. The dominant $680~\mathrm{K}$ anomaly coincides with the well-known tetragonal-to-orthorhombic transition, which manifests as the $a$--$b$ splitting observed in Fig.~\ref{lattice}.

For $\mathrm{La}_3\mathrm{Ni}_2\mathrm{O}_7$, we again measured a powder sample of the bilayer 2222 phase, shown in red in Fig.~\ref{DSC}(b). The structural anomaly identified near $560~\mathrm{K}$ in Fig.~\ref{lattice}(b) is accompanied by a sharp, fully reversible DSC peak, indicating an entropy change, with no significant changes upon thermal cycling. The enhanced signal amplitude partly reflects the more efficient thermal coupling achieved by completely filling the Al crucible with powder. The monolayer--trilayer 1313 variant, shown in black, exhibits a closely similar entropy release at the same temperature.

Remarkably, the trilayer compound $\mathrm{La}_4\mathrm{Ni}_3\mathrm{O}_{10}$ [Fig.~\ref{DSC}(c)] displays an entropy release at precisely the same temperature of $\sim560~\mathrm{K}$. In contrast, the three-dimensional perovskite end member $\mathrm{LaNiO}_3$ [Fig.~\ref{DSC}(d)] is featureless over the entire measured temperature range, as expected for this metallic Ni$^{3+}$ compound.

\subsection{Magnetism}
\begin{figure}
    \centering
    \includegraphics[width=1\linewidth]{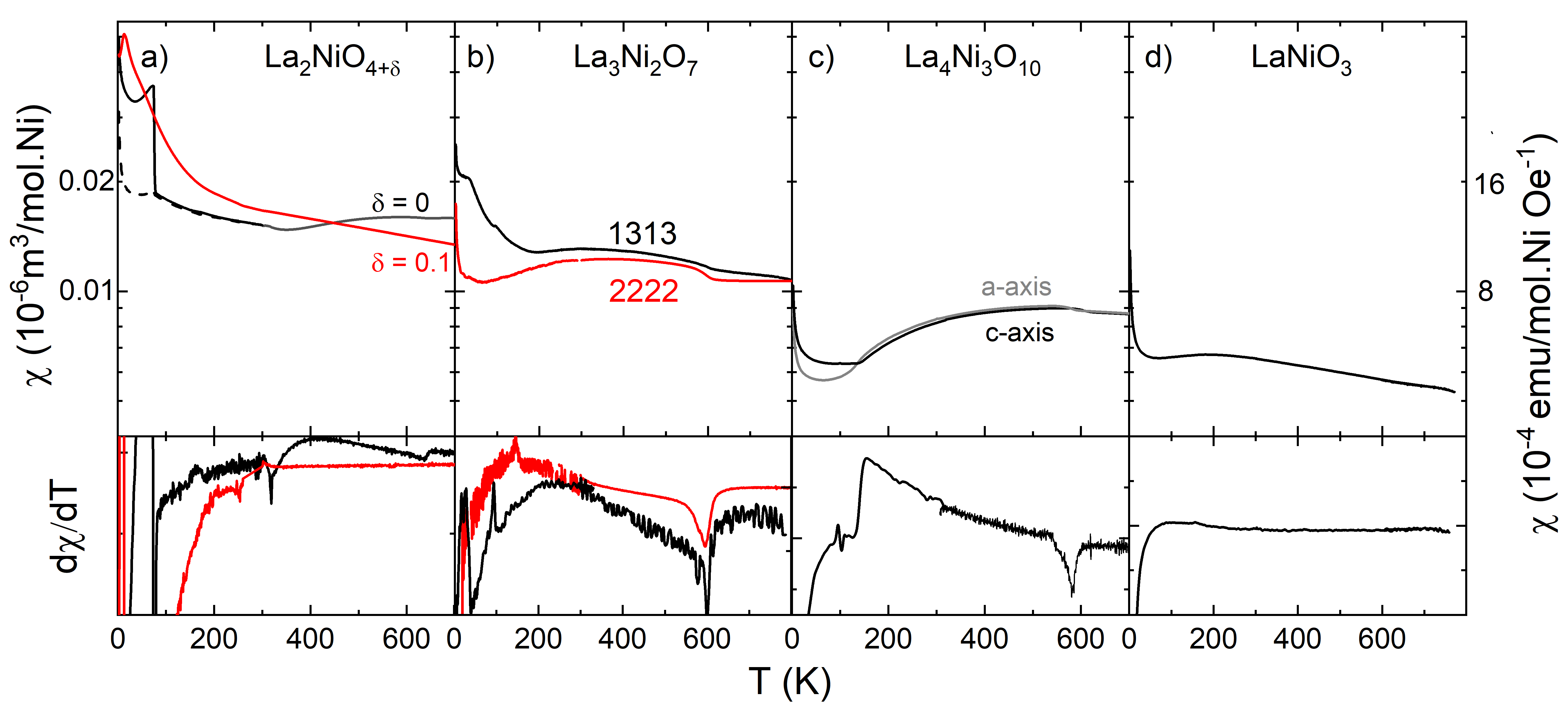}
    \caption{Magnetic susceptibility of single crystals of Ruddlesden--Popper $\mathrm{La}_{n+1}\mathrm{Ni}_n\mathrm{O}_{3n+1}$ for (a) $n=1$, (b) $n=2$, (c) $n=3$, and (d) $n=\infty$ in an external field of 0.1 T.}
    \label{Sus}
\end{figure}

To probe the magnetic response associated with these transitions, we performed SQUID magnetometry measurements over $2$--$380~\mathrm{K}$, complemented by high-temperature high-temperature measurements over $300-800~\mathrm{K}$ using the oven insert. The resulting data sets were combined and are shown in Fig.~\ref{Sus}. A shared $y$ axis is used to emphasize the systematic reduction of the susceptibility with increasing Ni oxidation state toward Ni$^{3+}$.  
For a $\mathrm{La}_2\mathrm{NiO}_4$ single crystal with a mass of $227~\mathrm{mg}$, the magnetic susceptibility exhibits a subtle increase near $670~\mathrm{K}$, followed by a broad maximum and a gradual decrease down to approximately $320~\mathrm{K}$. Below this temperature, the susceptibility increases weakly until a very sharp bifurcation between zero-field-cooled (ZFC) and field-cooled (FC) data appears at $80~\mathrm{K}$, superimposed on a Curie-like tail originating from dilute impurity spins. All of these magnetic anomalies correspond closely to the structural transitions identified in the lattice-parameter evolution.  

By contrast, for oxygen-intercalated $\mathrm{La}_2\mathrm{NiO}_{4+\delta}$ with $\delta=0.1$, the high-temperature transition is shifted to approximately $315~\mathrm{K}$, and the susceptibility follows an almost linear temperature dependence above this temperature. Upon cooling, a weak increase is observed until a sudden drop in the signal occurs near $18~\mathrm{K}$. The transition at $315~\mathrm{K}$ coincides with the full separation of the $a$ and $b$ lattice parameters, indicating a decoupling of the structural and magnetic transitions. Most importantly, this comparison demonstrates the dramatic impact of oxygen intercalation on the magnetic properties of layered nickelates.

Next, Fig.~\ref{Sus}(b) shows the magnetic response of $\mathrm{La}_3\mathrm{Ni}_2\mathrm{O}_7$, which has previously been reported for powder samples in Refs.~\cite{Sasaki1997,Kobayashi1996}. Consistent with those studies, we observe a pronounced increase in susceptibility near $590~\mathrm{K}$ in both a $6.7~\mathrm{mg}$ 1313 single crystal and a $5~\mathrm{mg}$ pellet of the 2222 phase. The magnitude of the maximum differs between the two polymorphs, which may indicate differences in effective exchange interactions, magnetic anisotropy, or oxygen stoichiometry. At lower temperatures, the behavior diverges more strongly: the 1313 polymorph exhibits a sharp increase near $200~\mathrm{K}$ and additional maxima at approximately $110~\mathrm{K}$ and $37~\mathrm{K}$. The features at $200~\mathrm{K}$ and $37~\mathrm{K}$ may originate from monolayer inclusions \cite{Khasanov2025c}. In contrast, the 2222 phase shows only a very weak anomaly near $145~\mathrm{K}$, which is discernible primarily in the temperature derivative of the susceptibility.  

For $\mathrm{La}_4\mathrm{Ni}_3\mathrm{O}_{10}$, pronounced anisotropy of the magnetic susceptibility along the crystallographic $c$ and $a$ axes has been reported previously \cite{Zhang2020M}. We therefore measured the full temperature-dependent anisotropy of a $3~\mathrm{mg}$ single crystal over the entire accessible temperature range. The anisotropy is strongly enhanced at the transition near $585~\mathrm{K}$, where the susceptibility increases in a manner similar to the $n=2$ compounds. Upon further cooling, the susceptibility passes through a broad maximum and then decreases until the magnetic transition centered at $139~\mathrm{K}$, where an additional increase of the signal is observed.  

Finally, for a $9~\mathrm{mg}$ single crystal of the perovskite end member $\mathrm{LaNiO}_3$, no comparable maxima or transitions are detected up to the highest measured temperatures. Instead, the susceptibility decreases monotonically with increasing temperature, consistent with Pauli paramagnetism in a correlated metal, with a broad maximum near $200~\mathrm{K}$ and a low-temperature Curie tail arising from impurity spins.  

Overall, Curie--Weiss fits over extended temperature ranges are not appropriate for the layered compounds, owing to the pronounced changes in susceptibility at high temperatures. Nevertheless, the absolute magnitude of the high-temperature susceptibility can be compared with the expected effective magnetic moments for the corresponding Ni oxidation states. For Ni$^{2+}$, $\mu_{\mathrm{eff}}\approx2.83\,\mu_B$, while for low-spin Ni$^{3+}$ the moment is reduced to $\mu_{\mathrm{eff}}\approx1.73\,\mu_B$. Accordingly, the expected moments are $2.83$ ($2.61$), $2.28$, $2.08$, and $1.73~\mu_B$ for $\mathrm{La}_2\mathrm{Ni}^{2+}\mathrm{O}_4$ (as grown $\mathrm{La}_2\mathrm{Ni}^{2.2+}\mathrm{O}_{4.1}$), $\mathrm{La}_3\mathrm{Ni}_2^{2.5+}\mathrm{O}_7$, $\mathrm{La}_4\mathrm{Ni}_3^{2.67+}\mathrm{O}_{10}$, and $\mathrm{LaNi}^{3+}\mathrm{O}_3$, respectively.  

Experimentally, we obtain high-temperature susceptibilities of $15.8$ ($12.8$), $10.7$, $8.5$, and $5.28 \times 10^{-9}~\mathrm{m}^3\,\mathrm{mol}^{-1}$, corresponding to effective moments of $2.83$ ($2.55$), $2.33$, $2.08$, and $1.64~\mu_B$, respectively. These values are in good agreement with expectations and hint at the nominal Ni oxidation states of the respective compounds.

\subsection{Electrical transport}

Finally, we turn to electrical transport measurements performed over the wide temperature range of $2$--$800~\mathrm{K}$. For these experiments, we carefully selected small, high-quality single crystals that were pre-characterized by single-crystal XRD. 

\begin{figure}
    \centering
    \includegraphics[width=1\linewidth]{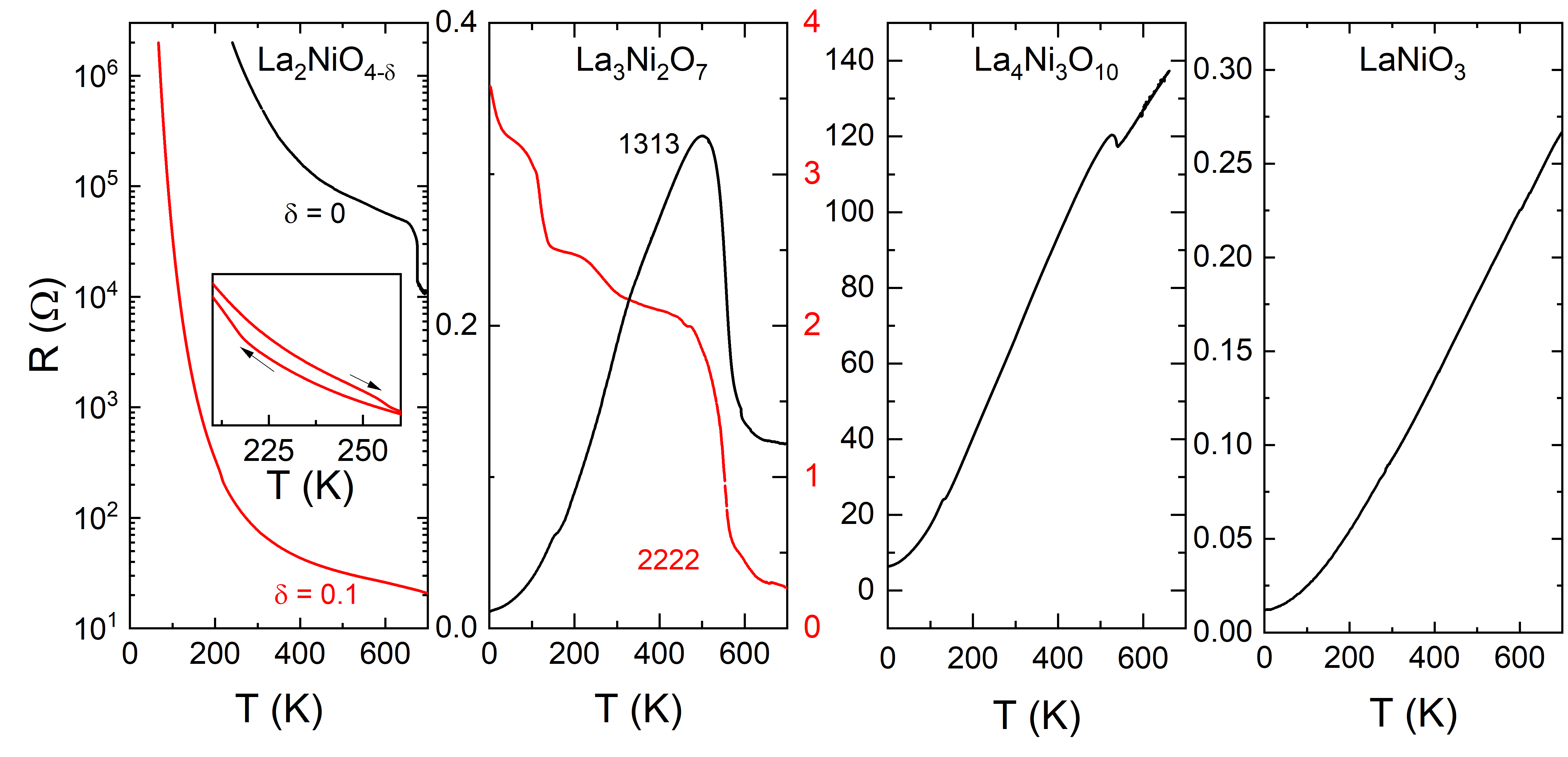}
    \caption{Electrical resistance of selected single crystals shown in the supplemental Materials for the Ruddlesden--Popper series $\mathrm{La}_{n+1}\mathrm{Ni}_n\mathrm{O}_{3n+1}$: (a) $n=1$ reduced (black) and as-grown (red), (b) $n=2$ 1313 (black) and 2222 (red), (c) $n=3$, and (d) $n=\infty$.}
    \label{R}
\end{figure}

Figure~\ref{R} summarizes the electrical resistance of the bulk-stable RP series. For the $n=1$ compound [Fig.~\ref{R}(a)], the reduced $\mathrm{La}_2\mathrm{NiO}_4$ crystal exhibits insulating behavior and shows a steep increase in resistance by approximately a factor of four at the expected tetragonal-to-orthorhombic transition near $680~\mathrm{K}$ (note the semilogarithmic scale), followed by an even stronger insulating temperature dependence. By contrast, the as-grown, slightly oxidized $\mathrm{La}_2\mathrm{NiO}_{4.1}$ sample displays a resistance reduced by roughly a factor of $500$ and only a weak anomaly near the now much lower transition temperature. This anomaly is strongly hysteretic and first order, as highlighted in the inset of Fig.~\ref{R}(a).

A similarly strong manifestation of the high-temperature structural anomaly is observed in the transport properties of $\mathrm{La}_3\mathrm{Ni}_2\mathrm{O}_7$, as shown in Fig.~\ref{R}(b). Both the bilayer 2222 and the monolayer--trilayer 1313 polymorphs display a sharp increase in resistance by a factor of approximately $3$--$6$ at the transition. At low temperatures, however, the two variants differ markedly in their transport behavior. We emphasize that the measured resistance is sensitive to the out-of-plane ($c$-axis) contribution: despite our efforts to maintain an in-plane current path, a slight misorientation of the bilayer crystal leads to a crossover from metallic to insulating behavior. A systematic study of the pronounced electrical anisotropy in these compounds is beyond the scope of the present work and will be addressed in future investigations.

Around $500~\mathrm{K}$, both the 2222 and 1313 phases exhibit an additional anomaly. While the 1313 polymorph crosses over into a metallic temperature dependence, the 2222 phase shows only a weak hump. At lower temperatures, the 1313 crystal exhibits a small peak reminiscent of features often attributed to density-wave formation in nickelates near $160~\mathrm{K}$, consistent with the entropy release observed in the specific-heat data (Fig.~\ref{Cp}). The bilayer crystal, on the other hand, displays a sharp increase in resistance near $146~\mathrm{K}$; here, however, the enhanced $c$-axis contribution must be taken into account. The additional anomalies near $38~\mathrm{K}$ and $200~\mathrm{K}$ are most likely associated with a minor monolayer impurity fraction.

The resistance of $\mathrm{La}_4\mathrm{Ni}_3\mathrm{O}_{10}$, shown in Fig.~\ref{R}(c), exhibits predominantly metallic behavior over the full temperature range. Only weak anomalies are observed, namely a small hump near $540~\mathrm{K}$ and another subtle feature near $135~\mathrm{K}$. Notably, the low-temperature density-wave transition produces only a minor change in resistance compared with the much more pronounced high-temperature anomaly.

Finally, the resistance of the perovskite end member $\mathrm{LaNiO}_3$, displayed in Fig.~\ref{R}(d), shows the expected metallic behavior throughout the entire investigated temperature range, with no evidence for additional phase transitions.

\subsection{Structure}

\begin{figure}
    \centering
    \includegraphics[width=1\linewidth]{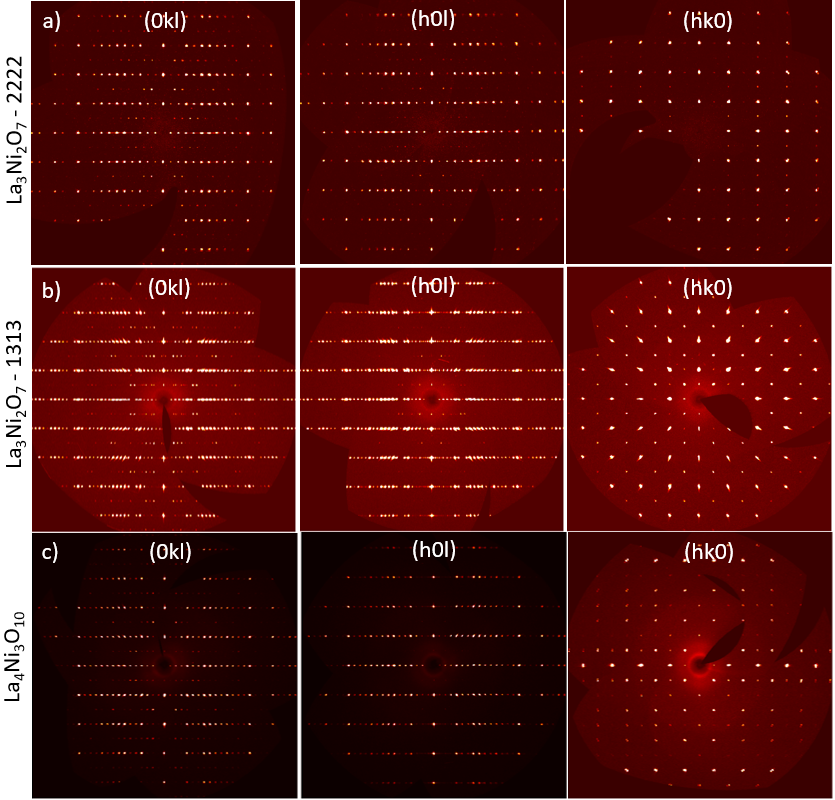}
    \caption{Room-temperature single-crystal X-ray diffraction integrated zonal maps of $\mathrm{La}_{n+1}\mathrm{Ni}_n\mathrm{O}_{3n+1}$ for (a) $n=2$ bilayer 2222, (b) $n=2$ monolayer--trilayer 1313, and (c) $n=3$. Shown are the $(0kl)$, $(h0l)$, and $(hk0)$ reciprocal-space sections.}
    \label{scXRD}
\end{figure}

Next, we turn to single-crystal X-ray diffraction (scXRD) measurements of $\mathrm{La}_{n+1}\mathrm{Ni}_n\mathrm{O}_{3n+1}$ over the temperature range $50$--$300~\mathrm{K}$. Figure~\ref{scXRD} presents integrated zonal maps for representative single crystals of bilayer $\mathrm{La}_3\mathrm{Ni}_2\mathrm{O}_7$--2222 (a), monolayer--trilayer $\mathrm{La}_3\mathrm{Ni}_2\mathrm{O}_7$--1313 (b), and trilayer $\mathrm{La}_4\mathrm{Ni}_3\mathrm{O}_{10}$ (c). In all cases, high-quality diffraction data were obtained using a conventional laboratory-based single-crystal XRD setup (as discussed in the methods section).  

For Fig.~\ref{scXRD}(a), corresponding to bilayer $\mathrm{La}_3\mathrm{Ni}_2\mathrm{O}_7$--2222, the established crystallographic description is space group \#63 $Amam$ (/$Cmcm$ which is equivalent with different axis settings) \cite{Liu2022growth}, where a recent study has proposed polar order with space group \#38 $Amm2$ \cite{Misawa2026}. From our refinements we obtain lattice parameters $a=5.4013(16)~\text{\AA}$, $b=5.4443(16)~\text{\AA}$, and $c=20.531(6)~\text{\AA}$, with the best structural solution with R$_1=2.54\%$ and wR$_1=6.87\%$ realized in space group \#50 $Ama2$ and a only slightly increased result with $Amm2$ (with R$_1=2.63\%$ and wR$_1=7.12\%$). Both are polar solutions and in case of $Amm2$ with the help of synchrotron data even charge order can be observed \cite{Misawa2026}. The obtained $R$ values of these polar variants are significantly lower than that achieved for refinements in $Amam$ (with R$_1=3.11\%$ and wR$_1=8.42\%$), indicating an improved description of the data. In reducing the symmetry from $Amam$ to $Ama2$ we loose centrosymmetry and the system realizes a polar structure similar to $Amm2$, the difference to $Amm2$ is that instead of mirroplanes we have a glide-plane. For $Amm2$ we find a significant Ni-O inplane bond distance difference of 1.962(9)$~\text{\AA}$ and 1.864(9)$~\text{\AA}$ between neighboring Ni atoms in plane, indeed indicating clear charge order. Combined with our observation of increased resistance we conclude that $Amm2$ is the realized solution. We attempted to refine interstitial oxygen occupancy but found no residual electron density on the fluorite interstitial sites. Instead, when allowing the apical oxygen occupancies to vary, we identify a very subtle vacancy on one of the two apical oxygen sites, resulting in a revised stoichiometry of $\mathrm{La}_3\mathrm{Ni}_2\mathrm{O}_{6.97(3)}$. By collecting data at five distinct temperatures and refining the oxygen occupancy independently for each data set, we obtain a statistical estimate of this oxygen deficiency and associated error bars (for crystallographic CIFs see \cite{Note1}).  

We next consider the data in Fig.~\ref{scXRD}(b), which correspond to the monolayer--trilayer $\mathrm{La}_3\mathrm{Ni}_2\mathrm{O}_7$--1313 polymorph. The reported crystallographic descriptions for this phase are not yet fully settled: two studies have refined the structure in space group \#65 $Cmmm$ \cite{Chen2024,Wang2024a} notably without any octahedral buckling, whereas our earlier high-pressure synchrotron diffraction study favored space group \#69 $Fmmm$ \cite{Puphal2024}. The latter discrepancy may arise from pronounced stacking faults, analogous to those discussed below for orthorhombic $\mathrm{La}_4\mathrm{Ni}_3\mathrm{O}_{10}$, or from variations in oxygen stoichiometry. The crystal investigated here exhibits clear superstructure reflections, consistent with features noted previously in Ref.~\cite{Chen2024}, where space group \#74 $Imma$ was proposed. We obtain lattice parameters $a=5.4355(5)~\text{\AA}$, $b=5.4652(5)~\text{\AA}$, and $c=40.698(4)~\text{\AA}$, with the best refinement achieved in space group $Imma$ (for cifs see \cite{Note1}). As for the bilayer crystal, we refined the oxygen occupancies. In contrast to the 2222 case, all framework oxygen sites are fully occupied; instead, we identify partially occupied interstitial oxygen on the fluorite site, yielding a revised stoichiometry of $\mathrm{La}_3\mathrm{Ni}_2\mathrm{O}_{7.15(2)}$. The associated uncertainty is again derived from independent refinements of data collected at multiple temperatures. This finding suggests that the $Imma$ superstructure is stabilized by oxygen intercalation, analogous to the well-known case of $\mathrm{La}_2\mathrm{NiO}_{4+\delta}$ \cite{Tamura1996,Huecker2004}. To our knowledge, this represents the first observation of a bulk single crystal of a higher-order RP phase exhibiting oxygen intercalation. As the crystal is obtained as grown without further annealing this hints at a slghtly higher oxidation state of the 1313 phase in comparison to 2222 and enables future seperation of the two phases in growth by precise temperature and pressure control. 

Local probes have already demonstrated that interstitial oxygen can significantly influence superconductivity in bilayer nickelate powders \cite{Dong2025}, and its relevance has also been highlighted in ozone-annealed thin films \cite{FerencSegedin2026}. Given the observation of $\delta\approx0.15$ excess oxygen, it is plausible that intermediate oxygen-ordered phases exist, similar to those known for $\mathrm{La}_2\mathrm{NiO}_{4+\delta}$ \cite{Tamura1996,Huecker2004}. In this context, the previously reported $Fmmm$ structures may correspond to crystals with  oxygen contents different from those refined in $Cmmm$.  

Turning to $\mathrm{La}_4\mathrm{Ni}_3\mathrm{O}_{10}$, the integrated zonal maps are shown in Fig.~\ref{scXRD}(c). Several structural descriptions have been reported for this compound: a monoclinic structure with space group \#14 $P2_1/a$ and an orthorhombic structure with space group \#64 $Bmab$ \cite{Zhang2020M,Yuan2024} and notably even the tetragonal $I4/mmm$ form stable at room temperature \cite{Shi2025tet}. Considering the comparatively high density-wave transition temperature above $150~\mathrm{K}$ and the frequent presence of $\mathrm{La}_3\mathrm{Ni}_2\mathrm{O}_7$ impurity reflections in powder XRD, the orthorhombic variant is likely associated with strong stacking disorder. Our investigated crystals unambiguously crystallize in the monoclinic $P2_1/a$ structure, with lattice parameters $a=5.4307(7)~\text{\AA}$, $b=5.4722(7)~\text{\AA}$, $c=14.2567(17)~\text{\AA}$, and $\beta=100.938(2)^\circ$. Consistent with the preceding cases, we refined the oxygen content and obtain a stoichiometry of $\mathrm{La}_4\mathrm{Ni}_3\mathrm{O}_{9.993(1)}$, with no detectable residual electron density on interstitial oxygen sites (for cifs see \cite{Note1}).

In the supplemental Material the low-temperature zonal maps are shown, enabling a direct comparison with the room-temperature maps in Fig. \ref{scXRD}. As evident, the zonal maps are remarkably similar, and no additional satellite reflections or other signatures of density-wave order are resolved within the sensitivity of laboratory XRD.


\begin{figure*}
    \centering
    \includegraphics[width=1\linewidth]{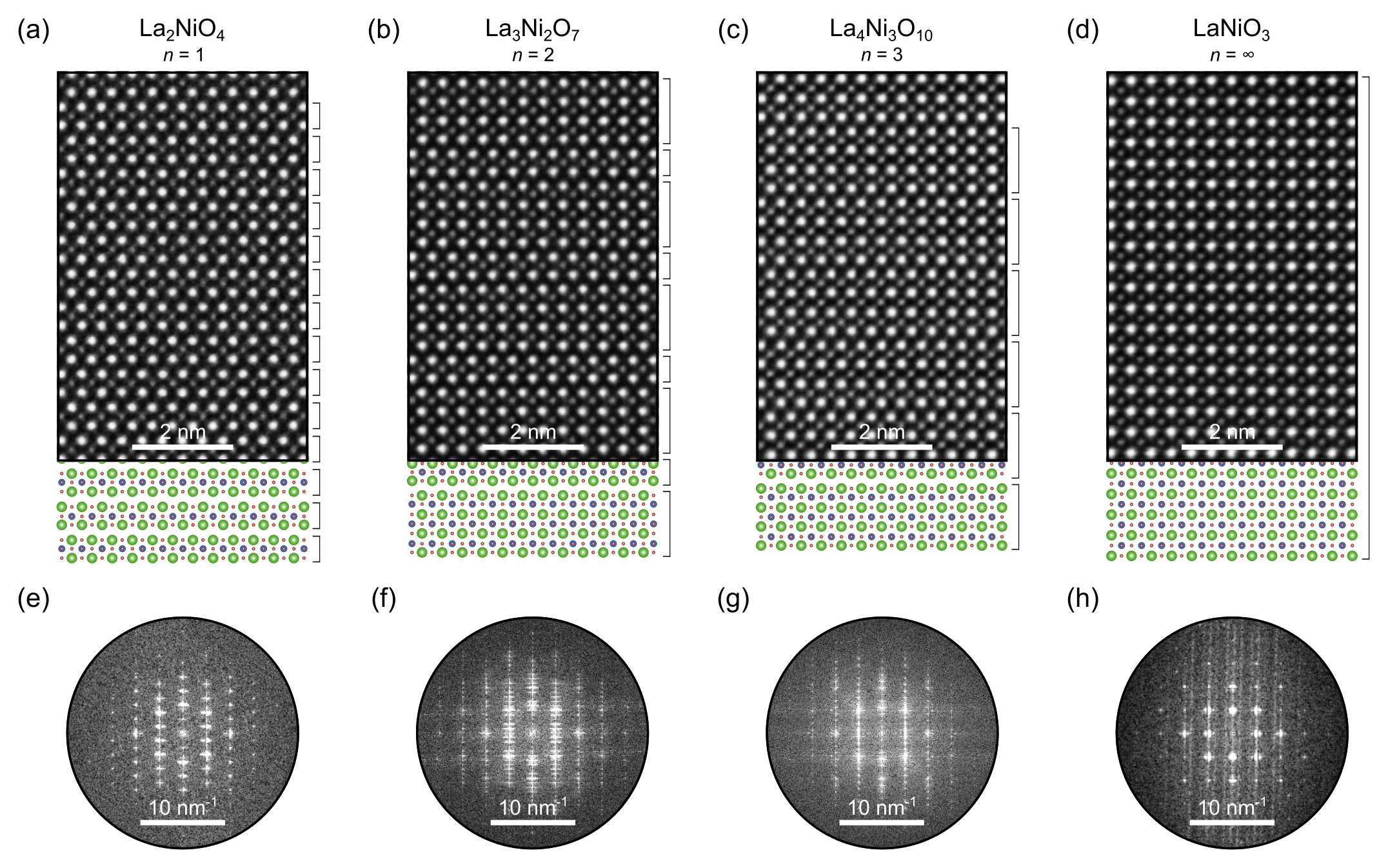}
    \caption{High-magnification STEM--HAADF images of the Ruddlesden--Popper $\mathrm{La}_{n+1}\mathrm{Ni}_n\mathrm{O}_{3n+1}$ series for (a) $n=1$, (b) $n=2$, (c) $n=3$, and (d) $n=\infty$. Brackets on the right highlight the characteristic stacking sequences. The corresponding structural models are overlaid at the bottom, with La shown in green, Ni in blue, and O in red (oxygen is not visible in HAADF contrast). The power spectrum for each structure is displayed under the corresponding figure, thus for (e) $n = 1$, (f) $n = 2$, (g) $n = 3$ and (h) $n = \infty$.}
    \label{STEM}
\end{figure*}

Figure~\ref{STEM} presents scanning transmission electron microscopy (STEM) high-angle annular dark-field (HAADF) images of representative regions from focused ion beam (FIB) lamellae prepared from single crystals extracted from the boules (see supplemental Materials). The brackets on the right indicate the distinct stacking sequences, and the atom columns are annotated with the corresponding structural motifs shown below each image. The atomic color code is La (green), Ni (blue), and O (red), noting that oxygen is not directly visible in HAADF imaging. Upon careful inspection, the stacking sequence can be readily identified from the lateral shifts of the La columns by half a lattice spacing.  

The images reveal extended stacking-fault-free regions corresponding to monolayer ($n=1$), alternating monolayer--trilayer ($n=2$, 1313), trilayer ($n=3$), and infinite ($n=\infty$) perovskite stackings. Together with the diffraction data, these observations support the formation of the intended RP structures through controlled stoichiometry and external oxygen pressure during growth. Importantly, although HAADF imaging is insensitive to oxygen, the absence of pronounced dislocations or severe distortions of the heavier cation sublattices indicates that no extreme octahedral tilts or structural instabilities are present on the local scale.
Dislocations and the average structure can be better observed via an indirect averaging. In Fig. \ref{STEM} (e-h) we show the power spectra corresponding to the fast Fourier transforms of the STEM images shown in Fig. \ref{STEM} (a-d), where now sharp spots correspond to spatial frequencies present in the atomic-resolution images. From the power spectra we can retrieve the lattice symmetry, and the positions of the spots give information about periodic lattice features. Nevertheless, power spectra are prone to streaking (here running vertically and horizontally) from finite image size effects, which we have ameliorated with gaussian apodization of the images. A comparison of these to our zonal diffraction images in Fig. \ref{scXRD} yields a qualitatively similar picture, where the observed lattice planes correspond to the unit cell symmetries inferred from the XRD data. Thus, we see an agreement of local nm sized (STEM) and averaged micron sized (XRD) structures and prove the phase purity and quality of the crystals revealing these high temperature transitions.

\section*{Conclusion}

In this work, we have presented a systematic investigation of the complete series of bulk-stable Ruddlesden--Popper nickelates $\mathrm{La}_{n+1}\mathrm{Ni}_n\mathrm{O}_{3n+1}$ with $n=1,2,3$, and $\infty$. 

Our powder X-ray diffraction studies suggest a previously underappreciated high-temperature structural transition near 560 K in the mixed-valence layered RP $n=2,3$ members distinct from the tetragonal to orthorhombic one. This transition is characterized by lattice-parameter changes. This transition was observed in both PXRD and DSC measurements for $n=2$ and $n=3$ compounds and manifests as a sharp, fully reversible entropy release. Its absence in the three-dimensional perovskite end member $\mathrm{LaNiO}_3$ hints that this transition might be an intrinsic feature of the mixed-valence layered RP topology. The widely discussed low-temperature density-wave transitions, by contrast, produce only comparatively subtle signatures in these measurements.

Similarly in Magnetic susceptibility measurements a pronounced high-temperature anomalies are observed explaining large magnetic anisotropies at room temperature. Electrical transport measurements demonstrate that high-temperature structural anomalies appear to have a direct and often dominant impact on transport, exceeding the signatures associated with low-temperature density-wave transitions. The sharp increase in resistance in La$_3$Ni$_2$O$_7$ combined with confirmation of a polar room temperature structure suggest a charge ordering scenario, while La$_4$Ni$_3$O$_{10}$ only has a minor bump similar to that observed in the density waves, possibly originating from the symmetry lowering from orthorhombic to its monoclinic structure that we find at room temperature.

For $\mathrm{La}_3\mathrm{Ni}_2\mathrm{O}_7$, we have shown that the bilayer 2222 and monolayer--trilayer 1313 polymorphs can host distinctly different oxygen stoichiometries possibly explaining the stabilization of a 1313 polymorph in OFZ. As grown, the 1313 phase reaches oxidation up to O$_{7.15}$, which stabilizes an $Imma$ superstructure. Here, we observed that while monolayer bucklings appear sensitive to oxygen intercalation, trilayer blocks remain buckled and preserve the buckling scheme. This suggests that higher-order RP nickelates can accommodate interstitial oxygen in bulk single-crystal form, similar to $\mathrm{La}_2\mathrm{NiO}_{4+\delta}$, and highlights oxygen content as a potentially important factor influencing their structural and electronic properties.

Future pressure-dependent studies of this high-temperature transition may shed further light on its possible relevance to superconductivity in RP nickelates.

\section*{Data availability}
The data are not publicly available but can be obtained from the corresponding author upon reasonable request.

\section*{Author contributions}
The manuscript was written, the samples were synthesized, Rietfeld- and, single crystal refinement, DSC, CP, and susceptibility were measured by P.P. The crystals were contacted by P.R. and transport was measured with P.P. J.N. measured single crystal diffraction. Low temperature and high temperature PXRD were measured by A.S. under the supervision of R.D.  STEM-HAADF were done by P.S.L. under the supervision of Y.E.S. and P.A.A. The project was initialized by B.K., H.T., M.I. and M.H.  All authors discussed the results and commented on the manuscript.

\section{Appendix: Methods}
\subsection{Crystal Growth}

$\mathrm{La}_{n+1}\mathrm{Ni}_n\mathrm{O}_{3n+1}$ ($n=1,2,3$ and $\infty$) single crystals were synthesized using the Optical Float Zone (OFZ) method. Before synthesis, La$_2$O$_3$ (Alfa Aesar, 99.9\%) was dried at 1100\degree C for 12~h, while NiO (Alfa Aesar, 99.0\%) was dried at 700\degree C for 6~h. The stoichiometric mixtures of La$_2$O$_3$ and NiO were ball-milled at 300~rpm for 6~h and then sintered twice in air at 1200\degree C ($n=1,\infty$), 1100\degree C ($n=2$), and 1150\degree C ($n=3$) for 12~h with ball milling as intermediate grinding. 
Cylindrically shaped feed and seed rods were prepared by ball-milling of the sintered materials, which were filled into rubber forms with 6~mm diameter. The rubber was evacuated and pressed in a stainless steel form filled with water using a Riken type S1-120 70 kN press. The extracted rods were transferred to alumina tubes and subsequently annealed at the same temperatures for 12~h in air. 

For $n=1$ Single crystal growth was carried out in an Optical Float Zone Furnace (FZ-T-10000-H-III-VPR) using four 1000 W halogen lamps. During growth, the feed and seed rods were counter-rotated at 24 rpm to minimize the diffusion zone near the solid–liquid interface. Argon flow of 100 cc~min$^{-1}$ was applied, and the growth rate stabilized at 4~mm~h$^{-1}$.\\

The single-crystal growth of higher $n$ was carried out in a high pressure, high-temperature OFZ furnace (model HKZ, SciDre GmbH, Dresden, Germany), that allows for gas pressures in the growth chamber up to 300 bar. The growth chamber (sapphire single crystal) has a length of 72~mm and a wall thickness of 20~mm. A Xe arc lamp operating at 5 kW was used operated at 4.7 kW power as a heating source within the vertical mirror alignment of the HKZ. The 14 cm feed and 4 cm seed rods were then aligned in the HKZ on steel holders followed by the installation of the high pressure chamber. Subsequently, for the growth of $n=2$ the chamber was pressurized with 15 bar oxygen gas and held at a flow rate of 0.1 l/min. After connecting the molten zone, the growth was carried out at a shutter opening around 33\% with moving the seed with a speed of 4 mm/h and the feed at 11 mm/h. Similarly a growth was carried out for $n=3$ with rods of according stoichiometry at 20 bar oxygen gas pressure. Finally, for LaNiO$_3$ with 80 bar gas pressure a slightly higher shutter opening was necessary. 

\subsection{X-ray Diffraction Measurements}
Variable temperature XRPD patterns were recorded on three machines:
1) Stoe STADI-P diffractometer operating in Debye–Scherrer geometry with monochromatic Ag K$_{\alpha _1}$ (Ge (111), $\lambda$ = 0.559 \AA) and a triple MYTHEN 1K detector setup, covering ~110$^\circ$ 2$\theta$ of each measurement. Temperature control was carried out with an FMB Oxford hot air gas blower, positioned perpendicular to the spinning 0.3 mm capillary with the analyzed sample. Powder patterns were recorded for 1 h each in 25 - 850 $^\circ$C temperature range with 50 $^\circ$C/step.
Owing to the extremely small size of bilayer 2222 single crystals and their frequent intergrowth with the monolayer--trilayer 1313 phase, we used phase-pure polycrystalline precursor rods to study the intrinsic structural evolution of the bilayer compound, while for other $n$ crushed crystals are used. Notably, the $\mathrm{La}_3\mathrm{Ni}_2\mathrm{O}_7$--1313 crystals contained a substantial fraction of the bilayer phase, which we refined separately. This independent refinement confirms the lattice-parameter evolution obtained from the powder measurements in crystals.

2) Rigaku SmartLab SE operating in a parallel beam reflection geometry with HyPix-3000 semiconductor detector in 1D mode.  The X-ray source was a 9 kW rotating anode tube with Cu K$_{\alpha _{1,2}}$ (Göbel mirror filtered, $\lambda$ = 1.54 \AA) radiation. The sample was packed into a single-crystal Si 911 sampleholder with 0.4 mm cavity. Before the measurement the sample was aligned in the Anton Paar HTK 1200N high-temperature chamber to account for a minimal 2$\theta$-offset at room temperature. XRPD patterns were recorded in 15 - 130$^\circ$ 2$\theta$ range with a scan speed of 2$^\circ$/min during the isothermal temperature steps every 20 $^\circ$C from room temperature until 700 $^\circ$C.
3) Bruker D8 Discover in Bragg-Brentano geometry with monochromatic Mo K$_{\alpha _{1}}$ ($\lambda$ = 0.71073 \AA) and 1D Lynxeye-2 detector. Sample cooling was acheived with a Oxford Cryosystems PheniX chamber. Each pattern was recorded for ~3 h in 2 - 70$^\circ$ 2$\theta$ range during isothermal temperature steps every 10 K/step down to 20 K, then at 15, 13 and again at 295 K.

Single-crystal data were collected on SMART APEXII CCD X-ray diffractometer, using graphite-monochromated Mo-K$_{\alpha}$ radiation ($\lambda=0.71073\,$\AA) and a D8-Venture CMOS-PhotonIV-C16 diffractometer, using multi-layer mirrored Mo-K$_{\alpha}$ radiation, generated by a microfocus I$\mu$S Diamond II sealed tube (Bruker AXS, Karlsruhe, Germany). Cooling was realized via a N-Helix low temperature device (Oxford Cryosystems, Oxford, United Kingdom). Pieces with lateral dimensions of approximately $\propto100$ $\mu$m were broken off from larger RP crystals. The pieces were mounted with some grease on a loop made of Kapton foil (Micromounts$^{TM}$, MiTeGen, Ithaca, NY). The reflection intensities were integrated with the SAINT subprogram in the Bruker Suite software package. The crystals showed systematic twinning by pseudo-merohedral twinning or reticular pseudo-merohedral twinning. To handle this a multi-scan absorption correction was applied either using SADABS or TWINABS. The structure was solved by direct methods and refined by full-matrix least-square fitting with the SHELXTL software package.

\subsection{STEM}
For the electron-transparent lamellae, a standard thinning procedure was carried out via focused ion beam (FIB) milling with Ga ions on a Thermo Fisher Scientific Scios 2 DualBeam FIB-SEM using the lift-out method. In the analyzed regions of the samples, the thicknesses were estimated below 40 nm via the Fourier log-ratio method.  High-resolution STEM imaging was performed employing a JEOL JEM-ARM200F STEM equipped with a probe Cs corrector (DCOR, CEOS GmbH) at an acceleration voltage of 200 kV, a convergence angle of 20.4 mrad (with a corresponding probe size of ~0.8 \AA) and a beam current smaller than 70 pA. Inner and outer collection semi-angles for STEM-HAADF images were of 75 and 310 mrad, respectively. Atomically-resolved images were acquired as multiple fast frames with a readout time of 0.8 $\mu$s/pixel, later rigidly aligned, summed and drift-corrected. The power spectra were computed for apodized (smooth zero-tapering at the edges) atomically-resolved images 

\subsection{PPMS: Specific Heat Measurements and electrical transport}
The specific-heat data were collected in the range of 2-300 K with the standard options of a Physical Property Measurement System (PPMS, Quantum Design) using a thermal relaxation method on cold pressed pellets of La$_3$Ni$_2$O$_{7}$ 2222 and single crystals of La$_2$NiO$_4$,  La$_3$Ni$_2$O$_7$, La$_4$Ni$_3$O$_{10}$, LaNiO$_{3}$. 
Similarly, electrical transport was measured using the standard PPMS option in the range of 2-340 K. All samples were contacted with  While for high temperature transport a glass cell was developed in-house that was installed in to a Nabertherm tube furnace, where we used the ac electrical transport method using a lock in and low fequencies f < 100 Hz. The temperature was read out via a Type-K thermocouple close to the sample.

\subsection{MPMS: Magnetic Susceptibility Measurements}
Magnetic susceptibility measurements were performed using a vibrating sample magnetometer (MPMS 3, Quantum Design) from 2-300 K and from 300-800 K using the oven option. In both cases the empty sample holders were measured and the corresponding background signal was subtracted.

\newpage
\bibliography{bib}

\section{Supplemental Material}
\subsection{Growth}

\begin{figure*}
    \centering
    \includegraphics[width=1\linewidth]{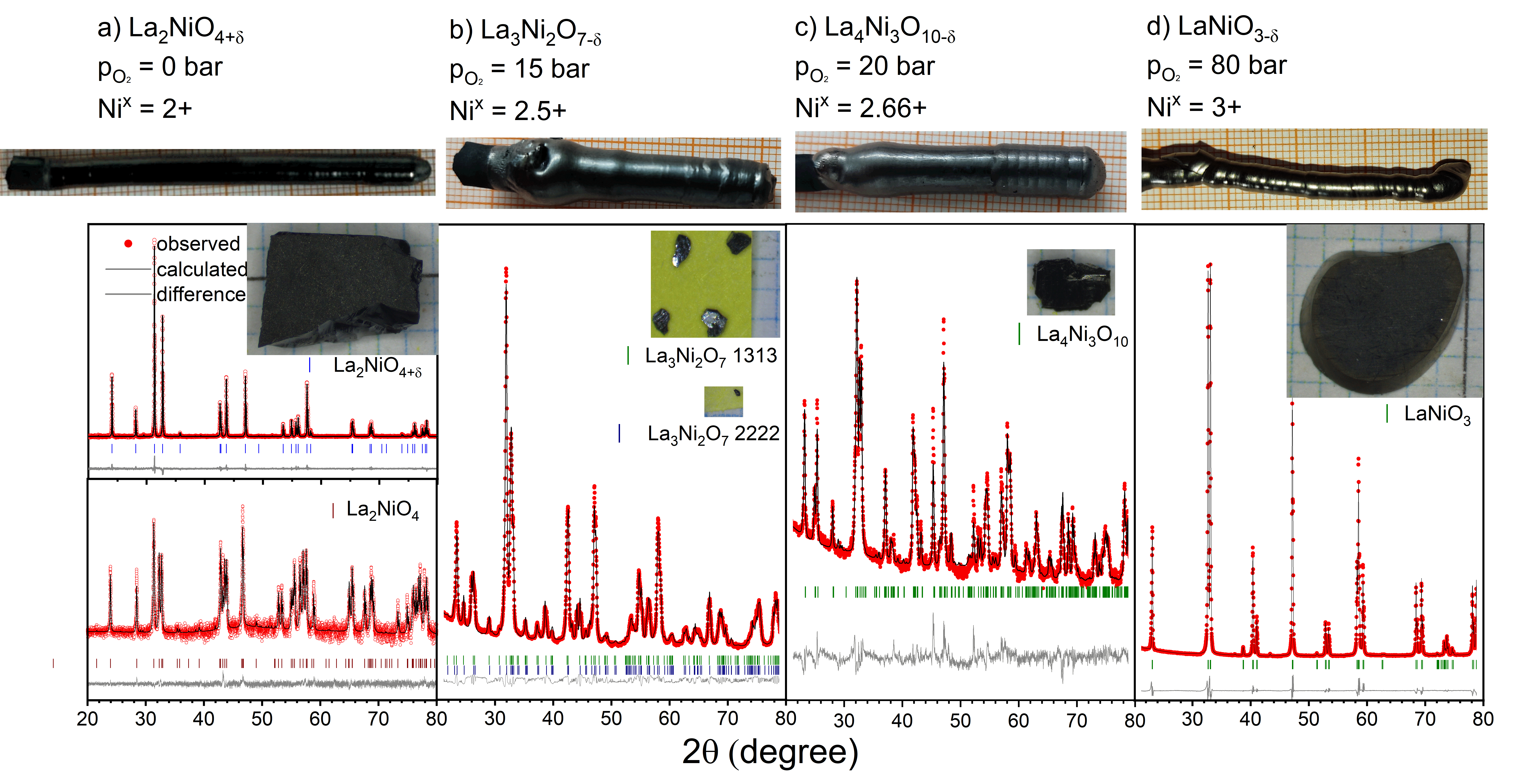}
    \caption{Optical floating-zone (OFZ) grown boules of nickelates labeled with the corresponding phase $\mathrm{La}_{n+1}\mathrm{Ni}_n\mathrm{O}_{3n+1}$, growth pressure, and nominal oxidation state, together with powder X-ray diffraction patterns obtained from crushed crystals: (a) $n=1$ (top panel: as-grown, bottom panel: reduced), (b) $n=2$, (c) $n=3$, and (d) $n=\infty$. Insets show photographs of representative extracted single crystals.}
    \label{phase}
\end{figure*}

For perovskite nickelates with $n=\infty$, $RE\mathrm{NiO}_3$, it is well established that the degree of octahedral buckling strongly depends on the ionic radius of the rare-earth ion. As the size of $RE$ decreases, the buckling increases, accompanied by a temperature-driven Mott metal--insulator transition and a symmetry lowering from $R\overline{3}c$ to $Pbnm$. Concomitantly, the transition temperature rises with increasing structural distortion.  

Perovskites ABO$_3$ exhibit remarkable chemical flexibility and are commonly realized with coordination numbers $\mathrm{CN}_A=12$ and $\mathrm{CN}_B=6$. The stability and distortion of the perovskite framework are often rationalized in terms of the Goldschmidt tolerance factor
$
t=\frac{r_A+r_O}{\sqrt{2}\,(r_B+r_O)},
$
introduced in Ref.~\cite{Goldschmidt1926}. An ideal cubic perovskite structure, such as that of $\mathrm{SrTiO}_3$, is obtained for $t\approx1$.

For $t<1$, increasing deviations from the ideal structure lead to progressively stronger octahedral tilting, evolving from rhombohedral structures such as $\mathrm{LaAlO}_3$ to orthorhombic distortions typified by $\mathrm{GdFeO}_3$. With increasing tilt amplitudes, the effective coordination number of the $A$-site cation is reduced. This change in coordination can also be viewed as a direct consequence of ionic radius ratios, consistent with Pauling’s rules, which relate preferred coordination numbers to relative ionic sizes \cite{Pauling1929}.  

In RP nickelates, the coordination number of the $A$-site cation is approximately $9$ in the vicinity of the rocksalt layers, while it reaches $12$ within the perovskite blocks in the absence of buckling.

The $n=2$ and $n=3$ RP phases are most commonly realized for divalent alkaline-earth $A$-site cations combined with tetravalent $B$-site cations, or in mixed-valence $A$-site systems that stabilize a trivalent $B$ site. A notable exception is nickel, as exemplified by the $\mathrm{La}_{n+1}\mathrm{Ni}_n\mathrm{O}_{3n+1}$ series, in which the formal oxidation state of Ni increases systematically with $n$ according to
$
\frac{3n-1}{n} \quad (2^+,\,2.5^+,\,2.67^+,\,\ldots),
$
a behavior that is, to date, unique among Ni-based RP phases. Consequently, both the evolving oxidation state of the $B$-site cation and the variation in the average $A$-site coordination number play a central role in governing the structural chemistry across the RP series. This unusual oxidation-state evolution likely underlies the stability of recently reported hybrid RP structures \cite{Puphal2024,Wang2024a,Chen2024,Li2024}.  

Importantly, the associated symmetry lowering and octahedral buckling are temperature dependent. As a result, these systems can realize both structural variants depicted in Fig.~\ref{struc}: the high-temperature tetragonal or cubic structures shown in the top row and the low-temperature buckled structures shown in the bottom row. Consistent with the evolution discussed above, the degree of buckling increases with increasing $n$, a trend highlighted in Fig.~\ref{struc} by the buckling angles listed beneath the corresponding space groups. The characteristic transition temperature can be estimated using the empirical relation proposed in Ref.~\cite{Herlihy2025}, which incorporates the concentration of Jahn--Teller-active species ($c_{\mathrm{JT}}$), the second moment (variance, $r\sigma^2$) of the $A$-site cation radius distribution, and the tolerance factor:
$
T_{\mathrm{LT}}(\mathrm{K}) \approx -31418\,(t - 0.89746) - 404.1\,c_{\mathrm{JT}} + 35236\,r\sigma^2.
$
Ref. \cite{Herlihy2025} estimates for La$_2$NiO$_4$ a $T_{LTO} = 780(39)$ K with $r\sigma^2 = 0$, $c_{\mathrm{JT}}=0$, however only when correcting the Ni radius for the high tempature value leading to t=0.873 with r$_{Ni}=0.716$ \AA. Considering the increase of Jahn Teller active specimen of Ni$^{3+}$ with $n$ a reduction of the $T_{LTO} \approx 566$ K for $n=2$ and to approximately 500 K for $n=3$ would be expected. Overall this formula shows an extreme sensitivity to the tolerance factor and this requires precise determination of the radius ratios.

To obtain the complete series of bulk-stable $\mathrm{La}_{n+1}\mathrm{Ni}_n\mathrm{O}_{3n+1}$ nickelates, we employed high-pressure optical floating-zone (OFZ) growth. In this technique, the effective oxygen partial pressure plays a decisive role in stabilizing a given RP phase, reflecting the systematic variation of the Ni oxidation state with $n$. As discussed in detail in the Methods section, OFZ growth was carried out at $0$~bar for feed rods with nominal stoichiometry $\mathrm{La}_2\mathrm{NiO}_4$, at $15$~bar for $\mathrm{La}_3\mathrm{Ni}_2\mathrm{O}_7$, at $20$~bar using $\mathrm{La}_4\mathrm{Ni}_3\mathrm{O}_{10}$ rods, and at $80$~bar for the perovskite $\mathrm{LaNiO}_3$. The resulting boules are shown in Fig.~\ref{phase}(a--d), together with powder X-ray diffraction patterns obtained from crushed crystals. Rietveld refinements indicate that the grown rods are largely homogeneous and phase pure, with the notable exception of Fig.~\ref{phase}(b), corresponding to $\mathrm{La}_3\mathrm{Ni}_2\mathrm{O}_7$.  

For $\mathrm{La}_3\mathrm{Ni}_2\mathrm{O}_7$, polymorphism is now well established in OFZ-grown samples, and both the bilayer (``2222'') phase and the monolayer--trilayer hybrid phase with ``1313'' stacking can be stabilized \cite{Puphal2024,Chen2024,Wang2024a}. In our boules, the 1313 phase constitutes the majority, with a minority fraction of the 2222 phase. Single crystals of the pure 2222 polymorph can only be isolated by mechanically fragmenting the material down to typical dimensions of $50\times50\times10~\mu\mathrm{m}^3$. This is illustrated in the inset of Fig.~\ref{phase}, which shows optical microscopy images of representative single crystals extracted from the boules.  

For all compositions, limited oxygen diffusion toward the center of the boule has a pronounced effect, promoting local stoichiometry variations and phase intergrowths. This behavior is well documented for $\mathrm{LaNiO}_3$ \cite{Zheng2020}. Such stoichiometry gradients can effectively give rise to incongruent-melting-like behavior, resulting in impurity-rich regions at the beginning and end of the boules \cite{Khasanov2025c}.  

\subsection{PXRD}

\begin{figure}
    \centering
    \includegraphics[width=1\linewidth]{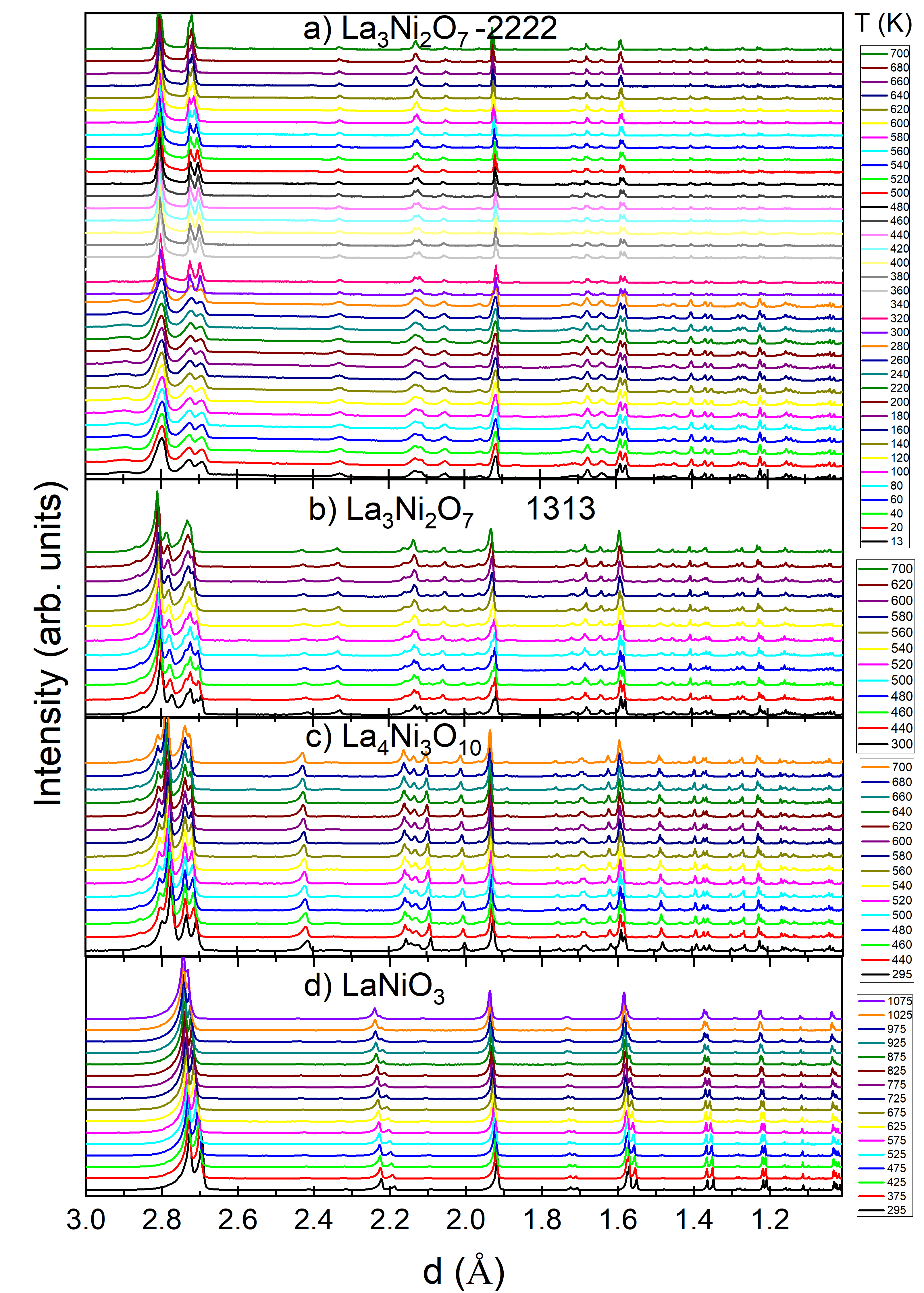}
    \caption{Powder X-ray diffraction results of (a) powder and (b--d) crushed single crystals of Ruddlesden--Popper $\mathrm{La}_{n+1}\mathrm{Ni}_n\mathrm{O}_{3n+1}$ for (a) $n=1$, (b) $n=2$, (c) $n=3$, and (d) $n=\infty$, measured at various temperatures (indicated in the legend) plotted versus the d-value due to varying x-ray sources.}
    \label{XRD}
\end{figure}

Representative sections of the diffraction patterns are shown in Fig.~\ref{XRD}. Diffraction data were collected over a typical $2\theta$ range of $10$--$100^\circ$ using different X-ray sources: Mo radiation for low-temperature measurements and Ag radiation for high-temperature capillary experiments, while the bilayer powder sample was measured using both Ag and Cu radiation. The measurement temperature is indicated in the legend in units of Kelvin and further details are given in the methods section. For all layered compounds, we consistently observe a progressive splitting of selected diffraction reflections upon cooling. In the bilayer case, measurements were extended down to $13~\mathrm{K}$, allowing us to track the separation of the $(200)$ and $(020)$ reflections directly. We performed individual Rietfeld refinement and their results are shown in the main manuscript.

\begin{figure}
    \centering
    \includegraphics[width=1\linewidth]{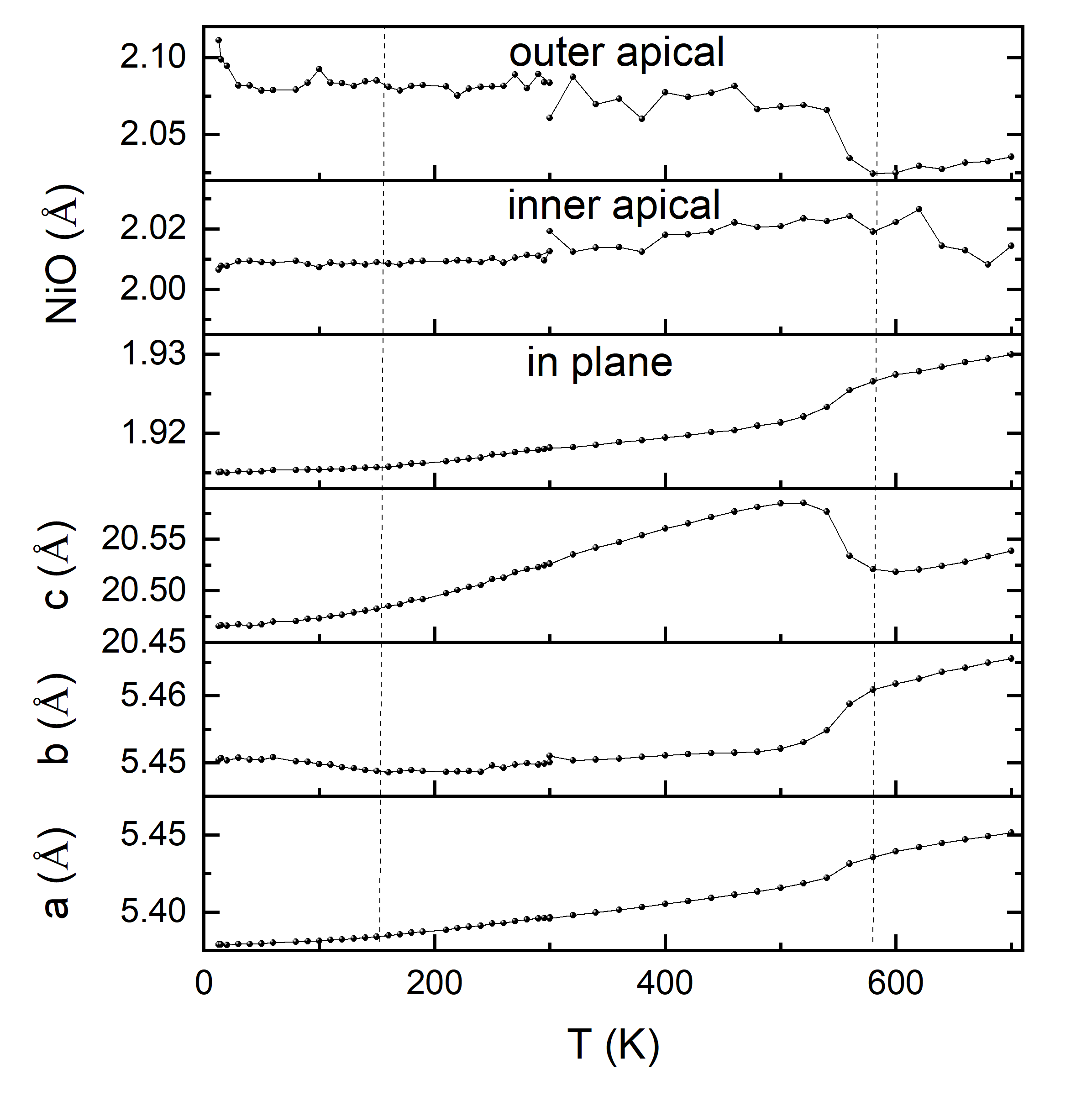}
    \caption{Temperature evolution of lattice parameters and Ni-O distances for Ruddlesden--Popper $\mathrm{La}_{3}\mathrm{Ni}_2\mathrm{O}_{7}$ using a $Fmmm$ cell.}
    \label{lattice2}
\end{figure}

The 560 K transition values further investigation and is most pronounced in the bilayer La$_3$Ni$_2$O$_7$. In Ref. \cite{Sasaki1997} the data was refined with a simple orthorhombic $Fmmm$ cell that does not account for octahedral bucklings, but they found an anomaly of oxygen positions around the transition. In Fig. \ref{lattice2} we show again the lattice constants for bilayer La$_3$Ni$_2$O$_7$ highlighting the two transitions with a dashed line now refined with this simple orthorhombic $Fmmm$ cell. As visible a jump of the apical oxygen position indicating enhanced octahedral distortions is evident from this refinement.

\begin{figure}
    \centering
    \includegraphics[width=1\linewidth]{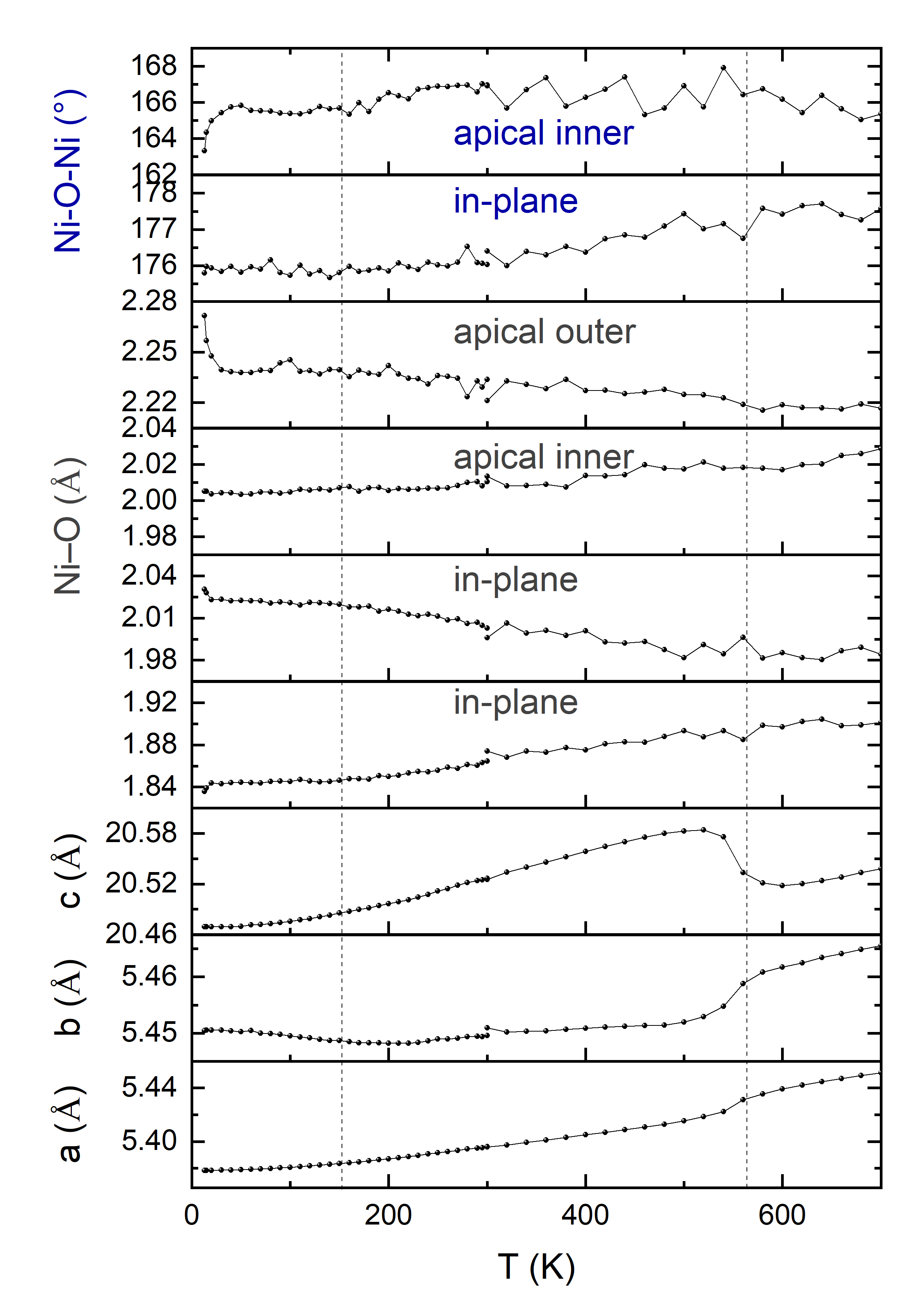}
    \caption{Temperature evolution of lattice parameters and Ni-O distances and Ni-O-Ni bond angles for Ruddlesden--Popper $\mathrm{La}_{3}\mathrm{Ni}_2\mathrm{O}_{7}$ using the currently established $Amam$ cell of the PXRD data.}
    \label{lattice3}
\end{figure}

However, in Fig. \ref{lattice3} we show the lattice constants, Ni-O distances and Ni-O-Ni bond angles when refining the same data with the orthorhombic $Amam$ cell, that allows for octahedral bucklings. Here, no clear anomaly besides that of the lattice constants can be seen in the Ni-O distances and the Ni-O-Ni bond angles. The transition appears more subtle indicating some trend changes. We should note despite the good quality of the data oxygen refinement with laboratory PXRD is to be considered with care and future neutron diffraction will shed light on the precise changes.

\subsection{Electrical contacts}

\begin{figure}
    \centering
    \includegraphics[width=1\linewidth]{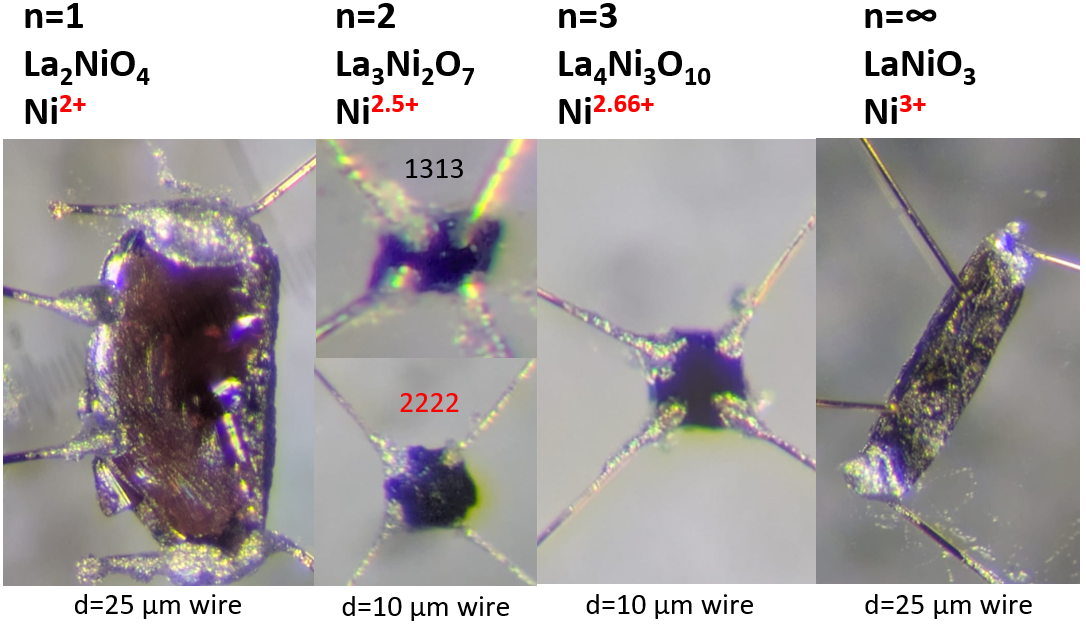}
    \caption{Contact geometries on selected single crystals of $\mathrm{La}_{n+1}\mathrm{Ni}_n\mathrm{O}_{3n+1}$ for (a) $n=1$, (b) $n=2$, (c) $n=3$, and (d) $n=\infty$.}
    \label{contacts}
\end{figure}
Establishing reliable electrical contacts on nickelates is considerably more challenging than on cuprates or conventional metals, because large contact resistances often dominate the measured response. We found that the use of high-temperature H20E Epo-Tek epoxy, cured for $5~\mathrm{min}$ at $350^\circ\mathrm{C}$, yields robust and low-resistance contacts. When applied inside a glove box, this procedure is even compatible with the highly metastable $\mathrm{La}_2\mathrm{NiO}_4$, which otherwise readily incorporates oxygen and changes from its characteristic brown transparent appearance, shown in Fig.~\ref{contacts}, to black. As illustrated in the figure, crystals of appropriate size were chosen to ensure phase-pure measurements, and special care was taken to avoid any unintended changes in oxidation state prior to contacting. All transport measurements presented here were performed upon heating over the temperature range from 2 to 800~K. This protocol was chosen because heating to 800~K appears to modify the oxygen content, as inferred from by a subtle but systematic increase in electrical resistance. Notably, differential scanning calorimetry (DSC) measurements (see Fig.~\ref{DSC}) demonstrate that the relevant transition persists despite partial oxygen loss. Consistent with this observation, we have also performed transport measurements after thermal cycling to high temperature, which yield qualitatively similar behavior (data not shown).

\subsection{Single crystal diffraction}
\begin{figure}
    \centering
    \includegraphics[width=1\linewidth]{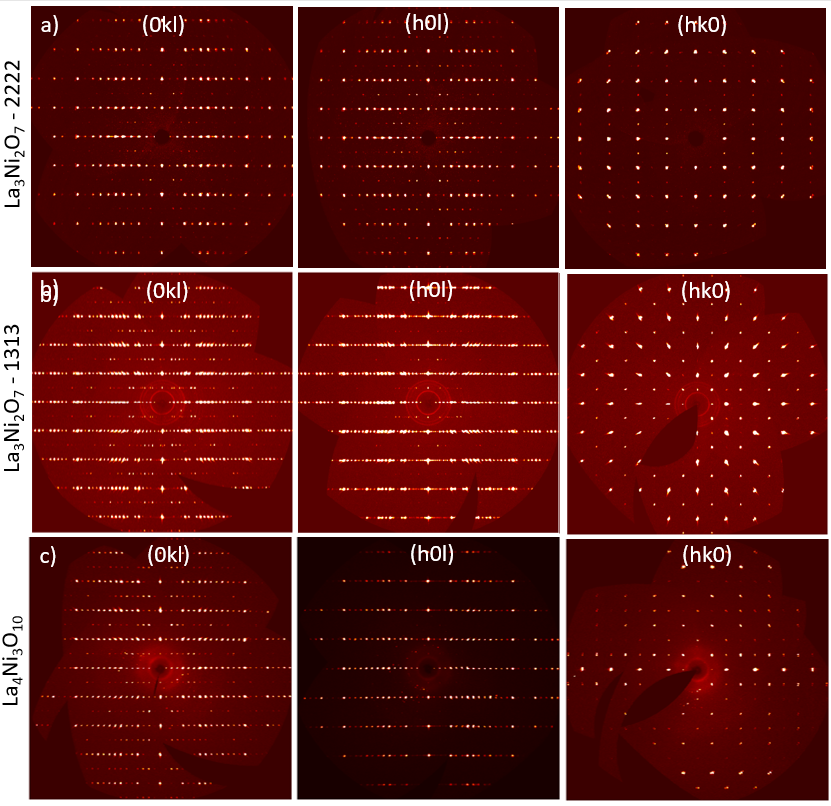}
    \caption{Low-temperature single-crystal X-ray diffraction integrated zonal maps of $\mathrm{La}_{n+1}\mathrm{Ni}_n\mathrm{O}_{3n+1}$ for (a) $n=2$ bilayer 2222 at 50 K, (b) $n=2$ monolayer--trilayer 1313 at 50 K, and (c) $n=3$ at 90 K. Shown are the $(0kl)$, $(h0l)$, and $(hk0)$ reciprocal-space sections.}
    \label{LTscXRD}
\end{figure}

Single-crystal X-ray diffraction (scXRD) was performed between 50 K and 380 K. Analysis of the data, including comparison of low-temperature (Fig. \ref{LTscXRD}) and higher-temperature (Fig. \ref{scXRD}) fourier maps, reveals no obvious major structural transitions within this range. This is indicated by the absence of additional superstructure satellites or significant intensity shifts.

\begin{figure}
    \centering
    \includegraphics[width=1\linewidth]{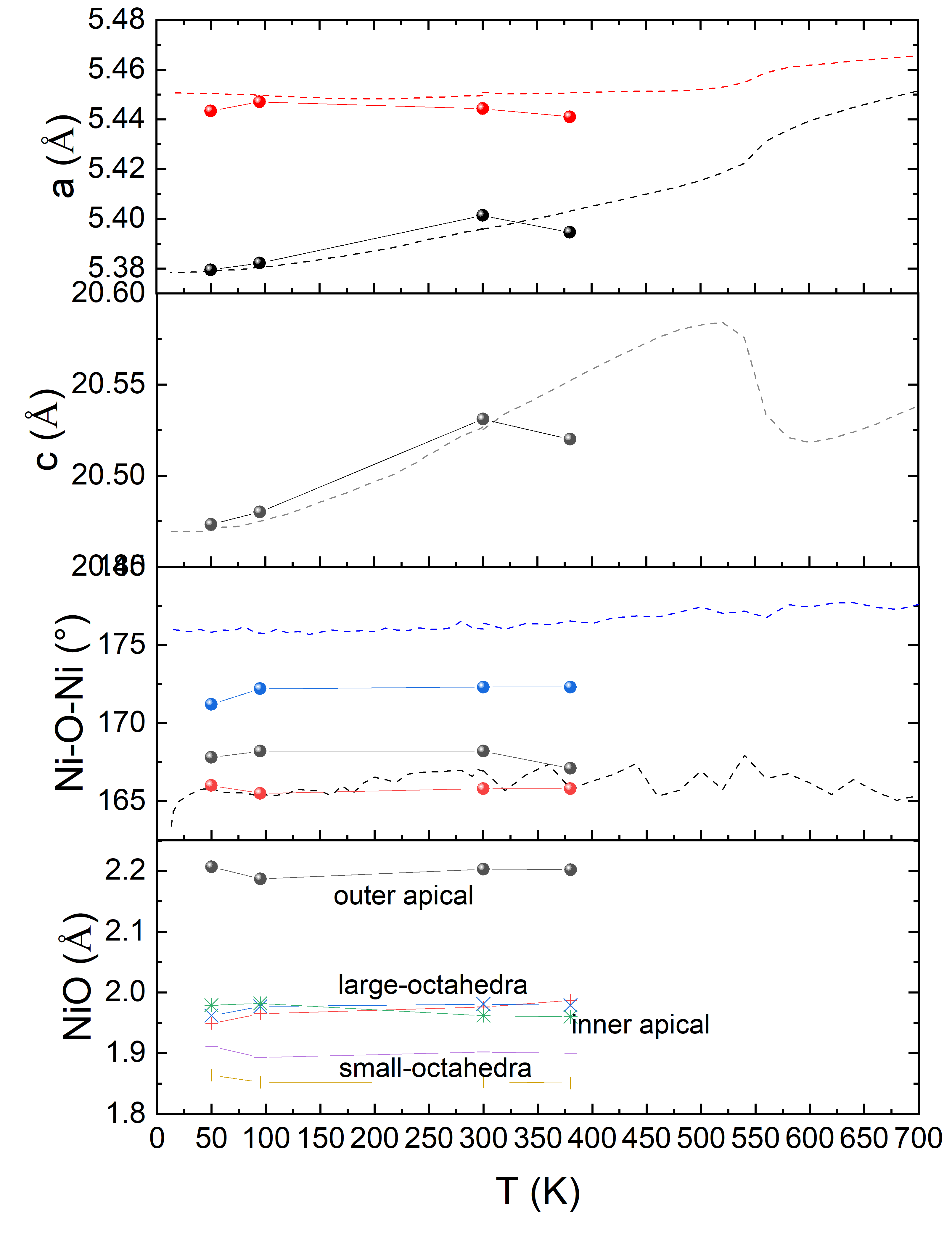}
    \caption{Temperature evolution of lattice parameters and Ni-O distances and Ni-O-Ni bond angles for Ruddlesden--Popper $\mathrm{La}_{3}\mathrm{Ni}_2\mathrm{O}_{7}$ in the $Amm2$ cell extracted from low-temperature single-crystal X-ray diffraction. The PXRD results from Fig. \ref{lattice3} are given as a dashed line to serve as a reference.}
    \label{lattice4}
\end{figure}

Analysis of single-crystal XRD data revealed more pronounced temperature dependencies in the lattice constants, Ni-O bond distances, and Ni-O-Ni bond angles compared to the crushed crystal PXRD data. In Fig. \ref{lattice4} we show the corresponding data in comparison with the powder data as a dashed line. Specifically, the limited single-crystal data for La$_3$Ni$_2$O$_7$-2222 suggests a strong volume change near the magnetic transition temperature of approximately 150 K. This volume change is reflected subtly in the bond distances and angles, which exhibit anomalous behavior. Most importantly throughout all temperatures, the charge ordered state is preserved with the smaller octahedral distance in-plane in average at 1.88(2)$~\text{\AA}$ and the larger octahedra with 1.97(2)$~\text{\AA}$.

\begin{figure*}
    \centering
    \includegraphics[width=1\linewidth]{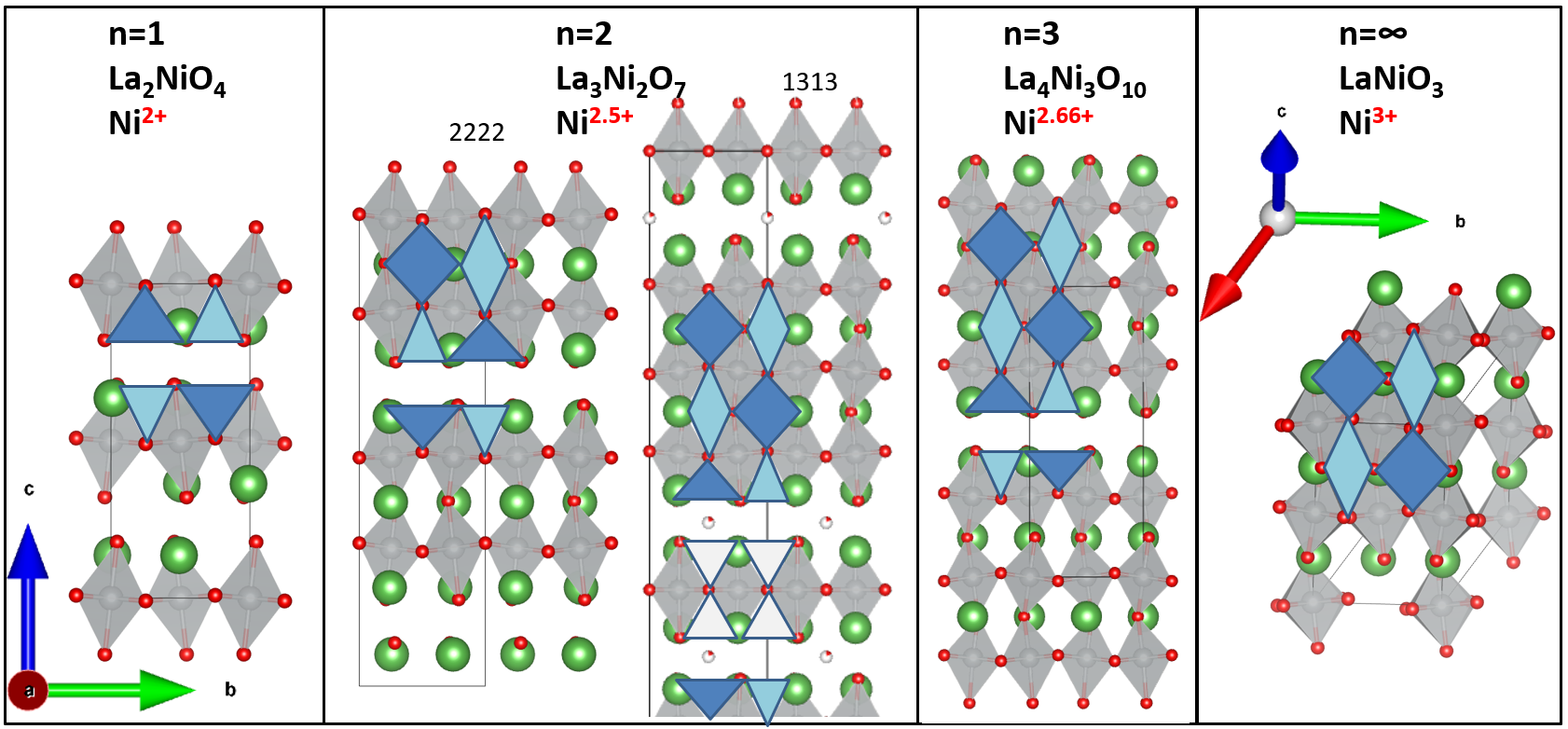}
    \caption{Structure at room temperature of $\mathrm{La}_{n+1}\mathrm{Ni}_n\mathrm{O}_{3n+1}$ obtained from single crystal XRD viewed along the diagonals of the octahedrons for $n=1$, $n=2$ bilayer 2222, and $n=2$ monolayer--trilayer 1313, $n=3$ and perovskite. The voids around the octahedrons are highlighted by dark blue, light blue and white to reveal the buckling principles.}
    \label{buckling}
\end{figure*}

To compare the bucklings in the entire family it is helpful to look at the structures along the diagonals as viewed in Fig. \ref{buckling}. Here, we highlight the voids in between the octahedrons via coloring. With the given tilting patterns in perovskites, we always find a regular octahedron (dark blue) alternating with a  tetragonally elongated octahedron (light blue). As visible in the right corner of Fig. \ref{buckling}. At the rocksalt layers however this octahedron pattern is interrupted and the octahedrons are shifted by one atom side (see Fig. \ref{STEM}). Along the diagonals this means that the void alters, i.e. the traingles at these intersection alter their color. This buckling pattern uniformally describes the room temperature structure of RP nickelates, with the excemption of the monolayer-trilayer hybrid. Here, both in our old $Fmmm$ solution \cite{Puphal2024} as well as in the here solved $Imma$ structure the monolayer blocks show no buckling (possibly from the intercalated oxygen, similar as it happens for pure monolayer La$_2$NiO$_{4.1}$). Surprisingly the buckling of the trilayer blocks is observed, while the $Cmmm$ solution, which likely is the solution without intercalation or subtle reduction, i.e. partial oxygen vacancies, shows no buckling. Inspecting the low temperature solution, we can see no major change of the overall buckling patterns, within resolution of our laboratory XRD, except for the lattice anomaly discussed above.

\end{document}